%% file: Hiertags_arxiv.tex
\documentclass[10pt]{article}

\usepackage{amsmath}
\usepackage{amssymb}
\usepackage{algpseudocode}
\usepackage{algorithmicx}
\usepackage{algorithm}

\usepackage{graphicx}

\usepackage{cite}

\usepackage{color}



\begin{document}

\begin{flushleft}
{\Large
\textbf{Extracting tag hierarchies}
}
\\
Gergely Tib{\'e}ly$^{1,3}$, 
P{\'e}ter Pollner$^{2,3}$,
Tam{\'a}s Vicsek$^{1,3}$ 
Gergely Palla$^{2,3,\ast}$
\\
{\bf $^1$ Dept.\ of Biological Physics, E{\"o}tv{\"o}s University,  Budapest, Hungary}
\\
{\bf $^2$ MTA-ELTE Statistical and Biological Physics Research Group of HAS, Budapest, Hungary}\\
{\bf $^3$ E{\"o}tv{\"o}s University, Regional Knowledge Centre, Sz{\'e}kesfeherv{\'a}r, Hungary}\\
$\ast$ E-mail: pallag@hal.elte.hu
\end{flushleft}

\section*{Abstract}
 Tagging items with descriptive annotations or keywords is a very natural way to compress and highlight information about the properties of the given entity. Over the years several methods have been proposed for extracting a hierarchy between the tags for systems with a "flat", egalitarian organization of the tags, which is very common when the tags correspond to free words given by numerous independent people. Here we present a complete framework for automated tag hierarchy extraction based on tag occurrence statistics. Along with proposing new algorithms, we are also introducing different quality measures enabling the detailed comparison of competing approaches from different aspects. Furthermore, we set up a synthetic, computer generated benchmark providing a versatile tool for testing, with a couple of tunable parameters capable of generating a wide range of test beds. Beside the computer generated input we also use real data in our studies, including a biological example with a pre-defined hierarchy between the tags. The encouraging similarity between the pre-defined and reconstructed hierarchy, as well as the seemingly meaningful hierarchies obtained for other real systems indicate that tag hierarchy extraction is a very promising direction for further research with a great potential for practical applications. 

Tags have become very prevalent nowadays in various online platforms ranging from blogs through scientific publications to protein databases. Furthermore, tagging systems dedicated for voluntary tagging of photos, films, books, etc. with free words are also becoming popular. The emerging large collections of tags associated with different objects are often referred to as folksonomies, highlighting their collaborative origin and the ``flat'' organization of the tags opposed to traditional hierarchical categorization. Adding a tag hierarchy corresponding to a given folksonomy can very effectively help narrowing or broadening the scope of search. Moreover, recommendation systems could also benefit from a tag hierarchy.

\section*{Introduction}
The appearance of tags in various online contents have become very common, e.g., tags indicate the topic of news-portal feeds and blog post, the genre of films or music records on file sharing portals, or the kind of goods offered in Web stores. By summarizing the most important properties of an entity in only a few words we ``compress'' information and provide a rough description of the given entity which can be processed very rapidly, (e.g., the user can decide whether the given post is of interest or not without actually reading it). The usage of tags, keywords, categories, etc., for helping the search and browsing amongst a large number of objects is a general idea that has been around for a long time in, e.g., scientific publications, library classification systems and biological classification. However, in the former examples the tagging (categorization) of the involved entities is hierarchical, with a set of narrower or broader categories building up a tree-like structure composed of ``is a subcategory of'' type relations. In contrast, the nature of tags appearing in online systems is rather different: they can usually correspond to any free word relevant to the tagged item, and they are almost never organized into a pre-defined hierarchy of categories and sub-categories \cite{Mika_folk_and_ont,Spyns_folk_and_ont,Voss_cond_mat}. Moreover, in some cases they originate from extensive collaboration as, e.g., in tagging systems like Flickr, CiteUlike or Delicious \cite{Cattuto_PNAS,Lambiotte_ct,Cattuto_PNAS2}, where unlimited number of users can tag photos, Web pages, etc., with free words. The arising set of free tags and associated objects are usually referred to as folksonomies, for emphasizing their collaborative nature. Since each tagging action is forming a new user-tag-object triple in these systems, their natural representation is given by tri-partite graphs, or in a more general framework by hypergraphs \cite{Lambiotte_ct,Newman_PRE,Caldarelli_PRE,Schoder_tags,Zhou_recommend_overview}, where the hyperedges connect more than two nodes together.

One of the very interesting challenges related to systems with free tagging is extracting a hierarchy between the appearing tags. Although most tagging systems are intrinsically egalitarian, the way users think about objects presumably has some built in hierarchy, e.g., ``poodle'' is usually considered as a special case of ``dog''. By revealing this sort of hierarchy from, e.g., tag co-occurrence statistics, we can significantly help broadening or narrowing the scope of search in the system, give recommendation about yet unvisited objects to the user \cite{Kazienko_chapter,Kazienko_paper}, or help the categorization of newly appearing objects. Beside the high relevance for practical applications, this problem is interesting also from the theoretical point of view, as marked by several alternative approaches proposed in the recent years. P.~Heymann and H.~Garcia-Molina introduced a tag hierarchy extracting algorithm based on analyzing node centralities in a co-occurrence network between the tags \cite{Garcia-Molina}, where connections between tags indicate the appearance of the tags on the same objects simultaneously and link weights correspond to the frequency of co-occurrences. Another interesting approach was outlined by  A.~Plangprasopchok and K.~Lerman \cite{Lerman_constr,Lerman_constr_2}, which can be applied to systems where users may define a shallow hierarchy for their own tags, and by agglomerating these shallow hierarchies we gain a global hierarchy between the tags. Further notable algorithms were given by P.~Schmitz \cite{Schmitz_constr}, using a probabilistic model and C.~Van~Damme et al. \cite{Van_Damme_constr}, integrating information from as many sources as possible.

In this paper we introduce a detailed framework for tag hierarchy extraction. Our intended main contributions to this field here are represented by the development of a synthetic, computer generated benchmark system, and the introduction of quality measures for extracted hierarchies. The basic idea of the benchmark system is to simulate the tagging of virtual objects with tags based on a pre-defined input hierarchy between the tags. When applying a hierarchy extraction algorithm to the generated data, the obtained tag hierarchy can be compared to the original tag hierarchy used in the simulation. By changing the parameters of the simulations we can test various properties of the tag hierarchy extracting algorithm in a controlled way. The different quality measures we introduce can be used to evaluate the results of a tag hierarchy extracting algorithm when the exact hierarchy between the tags is also known, (as, e.g., in case of the synthetic benchmark). Furthermore, we also develop new hierarchy extraction methods, which are competitive with the state of the art current methods.

These methods are tested on both the synthetic benchmark and on a couple of real systems as well. One of our data set contains proteins tagged with protein functions, where the extracted tag hierarchy can be compared to the protein function hierarchy of the Genome Ontology. The other real systems included in our study are given by tagged photos from the photo sharing platform Flickr and tagged movies from the Internet Movie Database (IMDb). In these cases, pre-defined ``exact'' tag hierarchies are not given, therefore, the outcomes of our hierarchy extraction algorithm can be evaluated only by visual inspection of smaller subgraphs in the obtained hierarchies. Luckily, as the tags correspond to English words in these systems, we can still get a good impression whether the obtained hierarchies are meaningful or not. 

Our tag hierarchy extraction methods are rooted in complex network theory. In the last 15 years the network approach has become an ubiquitous tool for analyzing complex systems \cite{Laci_revmod,Dorog_book}. Networks corresponding to realistic systems can be highly non-trivial, characterized by  a low average distance combined with a high average clustering coefficient \cite{Watts-Strogatz}, anomalous degree distributions \cite{Faloutsos,Laci_science} and an intricate modular structure \cite{GN-pnas,CPM_nature,Fortunato_report}. The appearance of node tags is very common in e.g., biological networks,\cite{Mason_nets_in_bio,Zhu_nets_in_bio,Aittokallio_nets_in_bio,Finocchiaro_cancer,Jonsson_Bioinformatics,Jonsson_BMC}, where they usually refer to the biological function of the units represented by the nodes (proteins, genes, etc.). Node features are also fundamental ingredients in the so-called co-evolving network models, where the evolution of the network topology affects the node properties and vice versa \cite{Eguiluz_coevolv,Watts_science,Newman_coevolv,Vazquez_cond_mat,Kozma_coevolv,Castellano_coevolv}. Meanwhile, hierarchical organization is yet another very relevant concept in network theory \cite{Laci_hier_scale,Sneppen_hier_measures,Newman_hier,Pumain_book,Sole_chaos_hier,Enys_hierarchy,Sole_cond_mat}. As networks provide a sort of ``backbone'' description for systems in biology, physics, chemistry, sociology, etc., whenever the related system is hierarchical, naturally, the given network is likely to preserve this aspect to some degree. This is supported by several recent studies, focusing on the dominant-subordinate hierarchy among crayfish \cite{Huber_crayfish}, the leader-follower network of pigeon flocks \cite{Tamas_pigeons}, the rhesus macaque kingdoms \cite{McCowan_macaque}, the structure of the transcriptional regulatory network of Escherichia coli \cite{Zeng_Ecoli}, and on a wide range of social \cite{Guimera_hier_soc,our_pref_coms,Sole_hier_soc} and technological networks \cite{Pumain_book}.

The two network based tag hierarchy extraction methods presented in this paper are both relying on the weighted network between the tags based on co-occurrence statistics. For the majority of the tags, the direct ancestor in the hierarchy is actually chosen from its neighbors in the network according to various delicate measures. 
\section*{Results}
\subsection*{Algorithms}
The reason for including both algorithm A and algorithm B in the paper is that algorithm A ``wins'' on the protein function data set, while algorithm B is better on the computer generated benchmarks and also seems to produce even more meaningful results in case of Flickr and IMDb. We made free implementation of both methods available at (http://hiertags.elte.hu).

\subsubsection*{Algorithm A}
The first stage corresponds to defining weighted links between the tags. Probably the most natural choice is given by the number of co-occurrences, (the number of objects tagged simultaneously by the given two tags). Since we are aiming at a directed network, (in which links are pointing from tags higher in the hierarchy towards descendants lower in the hierarchy), in this initial stage we actually assume two separate links pointing in the opposite direction for every pair of co-occurring tags, (with both links having the same weight). 

In the next step we prune the network by throwing away a part of the links. Instead applying a global threshold, for each tag $i$ we remove incoming links with a weight smaller than $\omega$ fraction of the weight of the strongest incoming link on $i$. According to our tests on the protein function data set, the quality of the results was only slightly effected by changing $\omega$. (Our quality measures and the description of the data sets are given in forthcoming sections). Nevertheless, an optimal plateau was observed in the quality as a function of $\omega$ between $\omega=0.3$ and $\omega=0.55$, as discussed in details Sect.S1.1.2 in the Supporting Information. Thus, in the rest of the paper we show results obtained at $\omega=0.4$. 

After the complete link removal process has been finished, the direct ancestor of tag $i$  is chosen from the remaining in-neighbors as follows. We calculate the $z$-score for the co-occurrence with each in-neighbor individually, given by the difference between the number of observed co-occurrences and the number of expected co-occurrences at random, scaled by the standard deviation, (based on the tag frequencies, more details on the $z$-score are given in Methods). The in-neighbor $j$ with the highest $z$-score is usually identified as the direct ancestor, and all other incoming links are deleted on $i$. However, there is a very important exception to this rule: in case the $i\rightarrow j$ link ``survived'' when thresholding the incoming links on $j$. This means that $i$ happens to be also a candidate for the ancestor of $j$, and actually the two tags are more likely to be siblings. In this scenario we go down the list of remaining in-neighbors of $i$ in the order of the $z$-score, until we find a candidate $l$ for which the link $i\rightarrow l$ was already deleted, and identify $l$ as the ancestor of $i$. In case no such in-neighbor can be found, $i$ becomes a local root, with temporally no incoming links. 

In the last phase of the algorithm we first choose a global root from the local ones according to the maximum entropy of their incoming link weight distribution: if the incoming link weights on $i$ are given by $w_{ij}$ with $\sum_jw_{ij}=W$, then entropy can be written as $-\sum_j \frac{w_{ij}}{W}\ln \frac{w_{ij}}{W}$. The reasoning behind this choice is that a large entropy usually corresponds to a large number of direct descendants with more or less uniform weight distribution. After the global root has been chosen, we go through the list of local roots in the order of their entropy, and link them under their partner with which they co-occur most frequently. (To avoid the formation of loops, we choose only from co-occurring partners  located in another subtree).

The result of the algorithm is a directed tree, since we assign one direct ancestor to every tag during the process, (except the global root), and we do not allow loops. The complexity of the algorithm can be estimated as $\mathcal{O}(Q)+\mathcal{O}(M \log M)$, where $Q$ denotes the number of objects, and $M$ stands for the number of links in the co-occurrence network between the tags. (The details and the  pseudo code of the algorithm are given in Sect.S1.1.1 in the Supporting Information).

\subsubsection*{Algorithm B}
In case of algorithm B the weight of the links in the network between the tags is the same as in algorithm A, namely the number of objects the tags co-occurred on. However, instead of parallel directed links pointing in the opposite direction, here we consider only single undirected links. Similarly to algorithm A, in the second phase we remove a part of the links from the network. However, in this case we use the $z$-score between connected pairs as a threshold, i.e., if the $z$-score is below 10, the given link is thrown away. (The optimal value for the $z$-score threshold was set based on experiments on our synthetic benchmark, as detailed in Sect.S1.2.2 in the Supporting Information.) There is one exception to the above rule of thresholding: if a tag appears on more than half of the objects of the other tag, then the corresponding link is kept even if the $z$-score is low. 

Next, the eigenvector centrality is calculated for the tags based on the weighted undirected network remaining after the thresholding, and the tags are sorted according to their centrality value. The hierarchy is built from bottom up: starting from the tag with the lowest eigenvector centrality we choose the direct ancestor of the given tag from its remaining neighbors according to a couple of simple rules. First of all, the ancestor must have a higher centrality. The reasoning behind this is that the eigenvector centrality is analogous to PageRank. Thus, the centrality of a tag is high if it is connected to many other high centrality tags, and therefore, higher centrality values are likely to appear on more frequent and more general tags. 

In case the tag $i$ has more than one remaining neighbor with a higher centrality value, we choose the candidate which is the most related to $i$ and the set of tags already classified as a descendant of $i$. This is implemented by aggregating the $z$-score between the given candidate and the tags in the branch starting from $i$, (including $i$ as well), and selecting according to the highest aggregated $z$-score value. We note that this is a unique feature of the algorithm: by aggregating over the descendants of $i$ we are using more information compared to simple similarity measures, and hence, are more likely to choose the most related candidate as the parent of $i$.

Since we iterate over the tags in reverse order according to their centrality value, and ancestors have always higher centralities compared to their descendants, no loops are formed during the procedure. The complexity of the method can be estimated as $\mathcal{O}(Q)+\mathcal{O}(N\cdot\ln N)$, where $Q$ stands for the number of objects and $N$ denotes the number of different tags. (The details and the pseudo code of the algorithm are given in Sect.S1.2.1 in the Supporting Information).

\subsection*{Measuring the quality of the extracted tag hierarchy} 
\subsubsection*{Simple quality measures}
Before actually discussing the results given by tag hierarchy extracting methods in different systems, we need to specify a couple of measures for quantifying the quality of the obtained hierarchies. The natural representation of a hierarchy is given by a directed acyclic graph (DAG), in which links are pointing from nodes at higher level in the hierarchy towards related other nodes lower in the hierarchy. If the exact tag hierarchy is known, the problem is mapped onto measuring the similarity between the DAG obtained from the tag hierarchy extraction method, the ``reconstructed'' graph, $\mathcal{G}_{\rm r}$ and the exact DAG, $\mathcal{G}_{\rm e}$. 

A simple and natural idea is taking the ratio of exactly matching links in $\mathcal{G}_{\rm r}$, denoted by $r_{\rm E}$, as a primary indicator. In case $\mathcal{G}_{\rm r}$ has only a single connected component, $r_{\rm E}$ is simply given by the number of links also present in $\mathcal{G}_{\rm e}$, divided by the total number of links in $\mathcal{G}_{\rm r}$, denoted by $M_{\rm r}$. However, if $\mathcal{G}_{\rm r}$ contains only a few links with a vast number of isolated nodes, this sort of normalization can lead to a unrealistically high $r_{\rm E}$ value, in case the links happen to be exactly matching. Thus, in the general case we normalize the number of exactly matching links by $\max(N-1,M_{\rm r})$, where $N-1$ corresponds to the number of links needed for creating a tree between the $N$ tags.

In a more tolerant approach we may also accept links between more distant ancestor descendant pairs according to the exact hierarchy, (e.g., links pointing from ``grandparents'' to ``grandchildren''). Beside the ratio of acceptable links, $r_{\rm A}$, we can measure the ratio of links between unrelated tags, $r_{\rm U}$ as well,  (these are pairs which are not connected by any directed path in $\mathcal{G}_{\rm e}$), and also the ratio of ``inverted'' links, $r_{\rm I}$, pointing in the opposite direction compared to $\mathcal{G}_{\rm e}$, or connecting more distant ancestor descendant pairs in the wrong direction. Furthermore, when $M_{\rm r}<N-1$, the ratio of missing links from $\mathcal{G}_{\rm r}$, denoted by $r_{\rm M}$, is another important indicator of the effectiveness of the algorithm. (If $\mathcal{G}_{\rm r}$ is composed of only a single component, $r_{\rm M}$ is 0 by definition.) Similarly to $r_{\rm E}$, all quality indicators introduced so far are normalized by $\max(N-1,M_{\rm r})$. These measures are not completely independent of each other, i.e., the ratio of acceptable links is always larger than or equal to the ratio of exactly matching links, $r_{\rm A}\geq r_{\rm E}$, and also $r_{\rm A}+r_{\rm I}+r_{\rm U}+r_{\rm M}=1$.

\subsubsection*{Normalized mutual information between hierarchies}
A somewhat more elaborate approach to measuring the quality of the reconstructed hierarchy can be given by the normalized mutual information, (NMI), introduced originally in information theory for measuring the mutual dependence of two random variables \cite{Kuncheva_mutinfo,Fred_mutinfo}. (The definition of the NMI in general is given in Methods). A very important application of the NMI is related to the problem of comparing different partitioning of the same graph into communities \cite{Danon_mutinfo,Lancichinetti_mutinfo}. The advantage of the NMI approach when comparing hierarchies is that the resulting similarity measure is sensitive not only to the amount of non-matching links, but also to the position of these links in the hierarchies. In other words, the change in the similarity is different for rewiring a link pointing to a leaf and for  rewiring a link higher in the hierarchy.

When judging the similarity between two hierarchies, a natural idea is to compare the sets of descendants for each tag in the corresponding DAGs. E.g., if the set of descendants of tag $i$ is $D_{\rm e}(i)$ in the exact hierarchy and $D_{\rm r}(i)$ in the reconstructed one, then the number of tags in the intersection of these two sets is given by $\left|D_{\rm e}(i)\cap D_{\rm r}(i)\right|$. Roughly speaking,  the higher the value of this quantity over all tags, the higher is the similarity between the two hierarchies. To  build a similarity measure from this concept in the spirit of the NMI, first we define $p_{\rm e}(i)=\left| D_{\rm e}(i)\right|/(N-1)$ as the probability for picking a tag from the descendants of $i$ at random in the exact hierarchy, where $N$ denotes the total number of tags in ${\mathcal G}_{\rm e}$. (Since the tag $i$ is not included in $D_{\rm e}(i)$, the possible maximum value for $\left|D_{\rm e}(i)\right|$ is $N-1$). Similarly, the probability for choosing a tag from the descendants of $i$ at random in ${\mathcal G}_{\rm r}$ is given by $p_{\rm r}(i)=\left| D_{\rm r}(i)\right|/(N-1)$, while the probability for picking a tag from the intersection between the descendants of $i$ in the two hierarchies can be written as $p_{\rm r,e}(i)=\left| D_{\rm e}(i)\cap D_{\rm r}(i)\right|/(N-1)$. Based on this, the NMI between the exact- and reconstructed hierarchies can be formulated as
\begin{equation}
I_{\rm e,r}=-\frac{2\sum\limits_{i=1}^Np_{\rm e,r}(i)\ln\left(\frac{p_{\rm e,r}(i)}{p_{\rm e}(i)p_{\rm r}(i)}\right)}{\sum\limits_{i=1}^Np_{\rm e}(i)\ln p_{\rm e}(i)+\sum\limits_{i=1}^Np_{\rm r}(i)\ln p_{\rm r}(i)}=\frac{2\sum\limits_{i=1}^N \left|D_{\rm e}(i)\cap D_{\rm r}(i)\right|\ln \left(\frac{\left|D_{\rm e}(i)\cap D_{\rm r}(i)\right| (N-1)}{\left|D_{\rm e}(i)\right|\cdot \left| D_{\rm r}(i)\right|}\right)}{
\sum\limits_{i=1}^N\left|D_{\rm e}(i)\right| 
\ln \left(\frac{\left|D_{\rm e}(i)\right|}{N-1}\right)+\sum\limits_{i=1}^N
 \left|D_{\rm r}(i)\right|\ln \left(\frac{\left|D_{\rm r}(i)\right|}{N-1}\right)} . \label{eq:NMI_DAG}
\end{equation}
This measure is 1 if and only ${\mathcal G}_{\rm e}$ and ${\mathcal G}_{\rm r}$ are identical, and is 0 if the intersections between the corresponding branches in the two hierarchies is of the same magnitude as we would expect at random, or in other words, if ${\mathcal G}_{\rm e}$ and ${\mathcal G}_{\rm r}$ are independent. The similarity defined in the above way is very closely related to the NMI used in community detection  \cite{Danon_mutinfo,Lancichinetti_mutinfo}, the analogy between the two quantities can be made explicit by an appropriate mapping from the hierarchy between the tags to a partitioning of the tags, (further details are given in Sect.S2.1 in the Supporting Information).  

\begin{figure}[!ht]
\begin{center}
\includegraphics[width=0.5\textwidth]{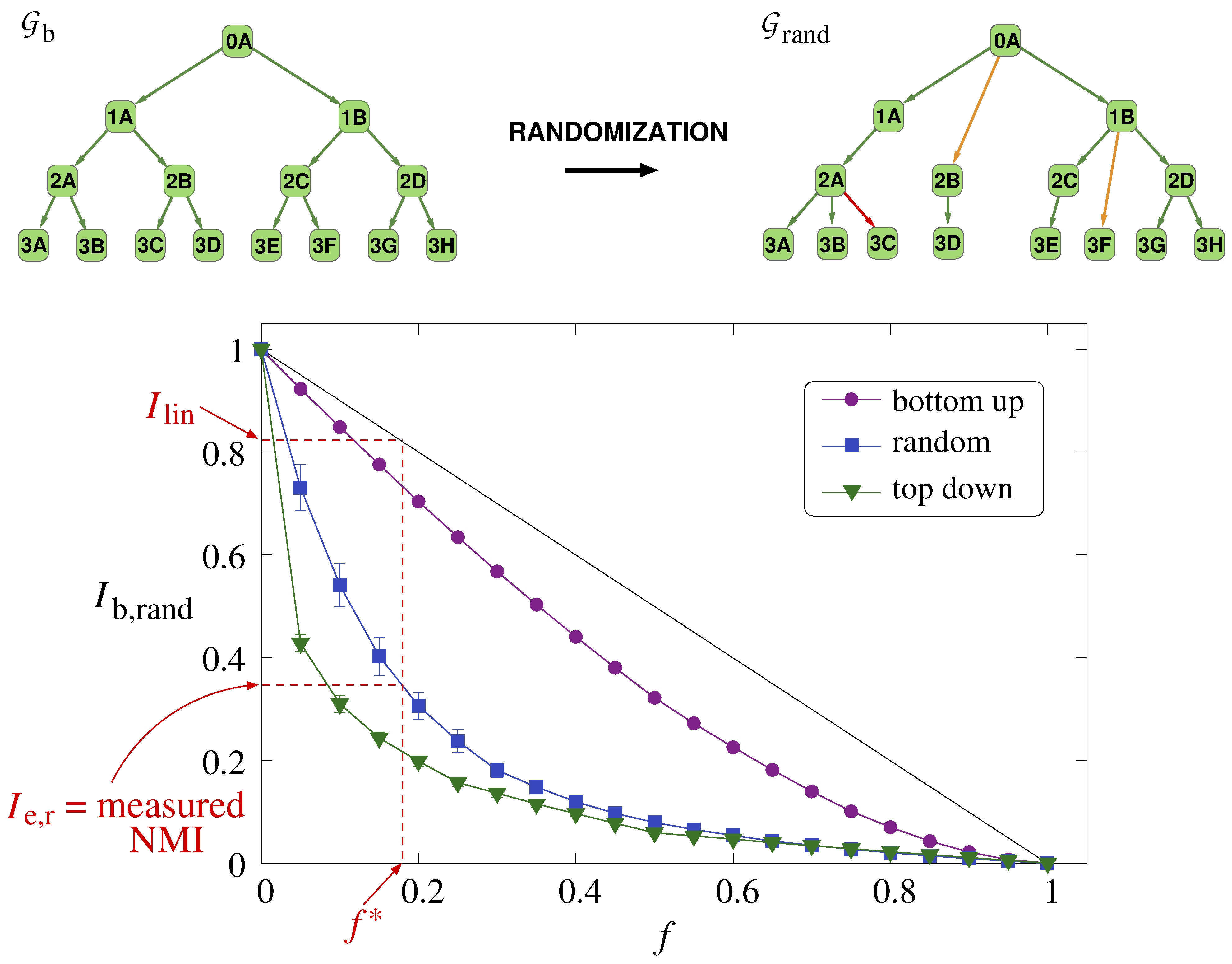}
\end{center}
\caption{
{\bf Using the normalized mutual information (NMI) for measuring the similarity between hierarchies.} We tested the behavior of the NMI by applying (\ref{eq:NMI_DAG}) to a binary tree of 1,023 nodes, ${\mathcal G}_{\rm b}$, and its randomized counter part, ${\mathcal G}_{\rm rand}$, obtained by rewiring the links at random, as shown in the illustration at the top. The decay of the obtained NMI is shown in the bottom panel as a function of the fraction of the rewired links, $f$. The three different curves correspond to rewiring the links in reverse order according to their position in the hierarchy (purple circles), rewiring in random order (blue squares) and rewiring in the order of the position in the hierarchy (green triangles). The concept of the linearized mutual information (LMI) for the general tag hierarchy reconstruction problem is illustrated in red: By projecting the measured $I_{\rm e,r}$ value onto the $f$ axis via the blue curve we obtain $f^*$, giving the fraction of rewired links in a randomization process with the same NMI value. The LMI is equal to $I_{\rm lin}=1-f^*$, corresponding to the fraction of unchanged links.  
\label{fig:NMI}}
\end{figure}

We examined the behavior of the NMI given in (\ref{eq:NMI_DAG}) by taking a binary tree of 1,023 nodes, $\mathcal{G}_{\rm b}$, and comparing it to its randomized counterpart, $\mathcal{G}_{\rm rand}$, obtained by rewiring a fraction of $f$ links to a random location. In Fig.\ref{fig:NMI}. we show the measured NMI as a function of $f$.  If we start the rewiring with links pointing to leafs, and continue according to the reverse order in the hierarchy, the NMI shows a close to linear decay as a function of $f$ almost in the entire $[0,1]$ interval (purple circles). However, if links are chosen in random order,  $I_{\rm b,rand}$ is decreasing much faster in the small $f$ region, with an overall non-linear $f$ dependency (blue squares). An even steeper decay can be observed when links are chosen in the order of their position in the hierarchy (green triangles). Nevertheless, $I_{\rm b,rand}\rightarrow 0$ when $f\rightarrow 1$ in all cases, thus, the similarity defined in this way is vanishing for a pair of independent DAGs. Meanwhile, the significant difference between the three curves displayed in Fig.\ref{fig:NMI}c shows that the NMI is sensitive also to the position of the rewired links in the hierarchy: rewiring the top levels of the hierarchy is accompanied by a drastic drop in the similarity, while changes at the bottom of the hierarchy cause only a minor decrease, which is linear in the fraction of rewired links.

This non-trivial feature of the NMI allows the introduction of another interesting quality measure for a reconstructed hierarchy. Supposing a similar randomization procedure on ${\mathcal G}_{\rm e}$ as shown in Fig.\ref{fig:NMI}, we may ask what fraction of links has to be rewired on average for reaching the same NMI as ${\mathcal G}_{\rm r}$? The formal definition of this measure is given as follows. Let $I(f)$ denote the average NMI obtained for a fraction of $f$ randomly rewired links, where the links are chosen in random order, $I(f)\equiv \left< I_{\rm e,rand}\right>_{f}$. By projecting the NMI between the exact- and reconstructed hierarchies, $I_{\rm e,r}$, to the $f$ axis using this function as
\begin{equation}
f^*= I^{-1}(I_{\rm e,r}),
\end{equation}
we receive the fraction of randomly chosen links to be rewired in ${\mathcal G}_{\rm e}$ for obtaining a randomized hierarchy with the same NMI as ${\mathcal G}_{\rm r}$, (see Fig.\ref{fig:NMI} for illustration). Based on that we define the linearized mutual information, (LMI) as 
\begin{equation}
I_{\rm lin}=1-f^*=1-I^{-1}(I_{\rm e,r}). \label{eq:LMI}
\end{equation}
This quality measure corresponds to the fraction of unchanged links in a random link rewiring process, resulting in a hierarchy with the same NMI as ${\mathcal  G}_{\rm r}$. (The reason for calling it ``linearized'' is that (\ref{eq:LMI}) is actually projecting $I_{\rm e,r}$ to the linear $1-f$ curve). By comparing the LMI to the fraction of exactly matching links, $r_{\rm e}$, we gain further information on the nature of the reconstructed DAG: If $I_{\rm lin}$ is significantly larger than $r_{\rm e}$, the reconstructed DAG is presumably better for the links high in the hierarchy, whereas if $I_{\rm lin}$ is significantly lower than $r_{\rm e}$, the reconstructed DAG is more precise for links close to the leafs. 

\subsection*{Real tagging systems}
\subsubsection*{Reconstructing the hierarchy of protein functions}
Although the primary targets of tag hierarchy extraction methods are given by tagging systems with no pre-defined hierarchy between the tags, for testing the quality of the extracted hierarchy we need input data for which the exact hierarchy is also given. A very important real tag hierarchy is provided by protein functions as described in the Gene Ontology \cite{GO}, organizing function annotations into three separate DAGs corresponding to ``biological process'', ``molecular function'' and  ``cellular component'' oriented description of proteins. The corresponding input data for a tag hierarchy extraction algorithm would be a collection of proteins, each tagged by its function annotations. Luckily, the Gene Ontology provides also a regularly updated large data set enlisting proteins and their known functions aggregated from a wide range of sources, (a more detailed description of the data set we used is given in Materials and Methods).

\begin{figure}[!ht]
\begin{center}
\includegraphics[width=0.9\textwidth]{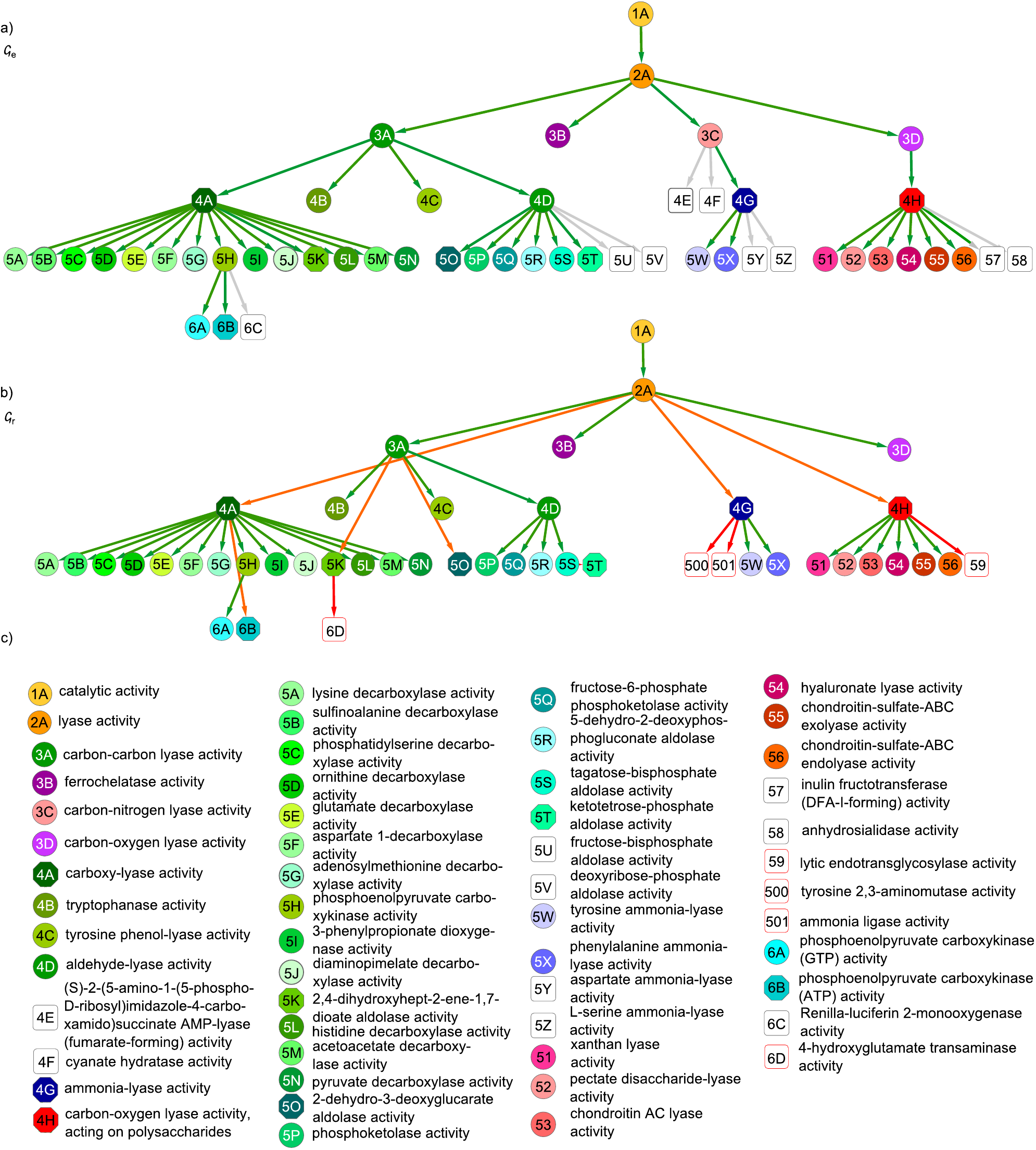}
\end{center}
\caption{
{\bf Comparison between the exact hierarchy and the reconstructed hierarchy obtained from algorithm A.} a) A subgraph in the hierarchy of protein functions, (describing molecular functions), according to the Gene Ontology, treated as the exact hierarchy, ${\mathcal G}_{\rm e}$. b) The hierarchy between the same tags obtained from running algorithm A on the tagged protein data set, (the reconstructed hierarchy, ${\mathcal G}_{\rm r}$). The exactly matching- and acceptable links are colored green and orange respectively, the unrelated links are shown in red, while the missing links are colored gray. c) The list of included protein functions in panels (a) and (b). 
\label{fig:GO_DAGs}
}
\end{figure}

In Fig.\ref{fig:GO_DAGs}a we show a smaller subgraph from the hierarchy between molecular functions given in the Gene Ontology, ${\mathcal G}_{\rm e}$, together with the subgraph between the same tags in the result obtained by running our algorithm A on the tagged protein data set, ${\mathcal G}_{\rm r}$, displayed in Fig.\ref{fig:GO_DAGs}b. The matching between the two subgraphs is very good: the majority of the connections are either exactly the same (shown in green), or acceptable (shown in orange), by-passing levels in the hierarchy and e.g., connecting ``grandchildren'' to ``grandparents''. The appearing few unrelated-- and missing links are colored red and gray, respectively. 

The quality measures obtained for the complete reconstructed hierarchy are given in table \ref{table:results}. For comparison we also evaluated the same measures for algorithm B, the algorithm by P.~Heymann and H.~Garcia-Molina, and the algorithm by P.~Schmitz. According to the results all 4 methods perform rather well, however, our algorithm seems to achieve the best scores. Although the ratio of exactly matching links is  $r_{\rm E}=21\%$, (which is not very high), the ratio of acceptable links is reaching $r_{\rm A}=66\%$, which is very promising. The NMI given by (\ref{eq:NMI_DAG}) is $I_{\rm e,r}=35\%$, however, the LMI according to (\ref{eq:LMI}) is $I_{\rm lin}=78\%$. (The corresponding plot showing the decay of the NMI between the Gene Ontology hierarchy and its randomized counterpart is given in Sect.S2.2 in the Supporting Information). Thus, the similarity between our reconstructed hierarchy and the hierarcy from the Gene Ontology is so high that if we would randomize the Gene Ontology, (by rewirnig the links in random order), the same NMI value would be reached already after rewiring 22\% of the links. The large difference between $I_{\rm lin}$ and $r_{\rm E}$ in favour of $I_{\rm lin}$ indicates that our algorithm is better at predicting links higher in the hierarchy. E.g., in a randomization with random link rewiring order keeping only $r_{\rm E}=21\%$ of the links unchanged, the NMI would be around $2\%$ instead of the actualy measured $I_{e,r}=35\%$. The reason why $I_{\rm e,r}$ can stay relatively high for the reconstructed hierarchy is that the majority of the non-matching links are low in the hierarchy, therefore, have a smaller effect on the NMI.

\begin{table}[!ht]
\caption{
\bf{Quality measures for the reconstructed hierarchies in case of the protein function data set}}
\begin{tabular}{|l|c|c|c|c|c|c|c|}
\hline
    & $r_{\rm E}$ & $r_{\rm A}$ & $r_{\rm I}$ & $r_{\rm U}$ & $r_{\rm M}$ & $I_{\rm e,r}$ & $I_{\rm lin}$
 \\ \hline  
algorithm A & 21\%   &66\% & 2\% & 32\%& 0\% & 35\% & 78\% \\
\hline
algorithm B &  20\% & 52\% & 3\% & 44\% & 1\% & 30\% & 75\% \\
\hline 
P.~Heymann \& H.~Garcia-Molina & 19\% & 51\% & 3\% & 46\%&  0\%& 30\% & 75\% \\
\hline
P.~Schmitz &  18\% & 65\% & 2\% & 23\% & 10\% & 30\% & 75\% \\
\hline
\end{tabular}
\begin{flushleft}
The quality of the tag hierarchy obtained for the tagged protein data set, ${\mathcal G}_{\rm r}$, was evaluated by comparing it to the hierarchy of protein functions in the Gene Ontology, ${\mathcal G}_{\rm e}$. The quality measures presented in the different columns are the following: the ratio of exactly matching links in ${\mathcal G}_{\rm r}$,denoted by $r_{\rm E}$, the  ratio of acceptable links, $r_{\rm A}$, (connecting more distant ancestor-descendant pairs), the ratio of inverted links, $r_{\rm I}$, (pointing in the opposite direction), the ratio of unrelated links, $r_{\rm U}$, (connecting tags on different branches in ${\mathcal G}_{\rm e}$), the ratio of missing links in ${\mathcal G}_{\rm e}$, denoted by $r_{\rm M}$, the normalized mutual information between the two hierarchies, $I_{\rm e,r}$, and the linearized mutual information, $I_{\rm lin}$, corresponding to the fraction of exactly matching links remaining after a random link rewiring process stopped at NMI value given by $I_{\rm e,r}$. The different rows correspond to results obtained from algorithm A (1$^{\rm st}$ row), algorithm B (2$^{\rm nd}$ row),the method by  P.~Heymann \& H.~Garcia-Molina (3$^{\rm d}$ row), and the algorithm by P.~Schmitz (4$^{\rm th}$ row).
\end{flushleft}
\label{table:results}
\end{table}

\subsubsection*{Hierarchy of Flickr tags}
One of the most widely known tagging systems is given by Flickr, an online photo management and sharing application, where users can tag the uploaded photos with free words. Since the tags are not organized into a global hierarchy, this system provides an essential example for the application field of tag hierarchy extracting algorithms. We have run our algorithm B on a relatively large, filtered sample of photos, (the details of the construction of our data set are given in Methods). Although the ``exact'' hierarchy between the tags is not known in this case, since the tags correspond to English words, we can still give a qualitative evaluation of the  result just by looking at smaller subgraphs in the extracted hierarchy. 

\begin{figure}[!ht]
\begin{center}
\includegraphics[width=0.9\textwidth]{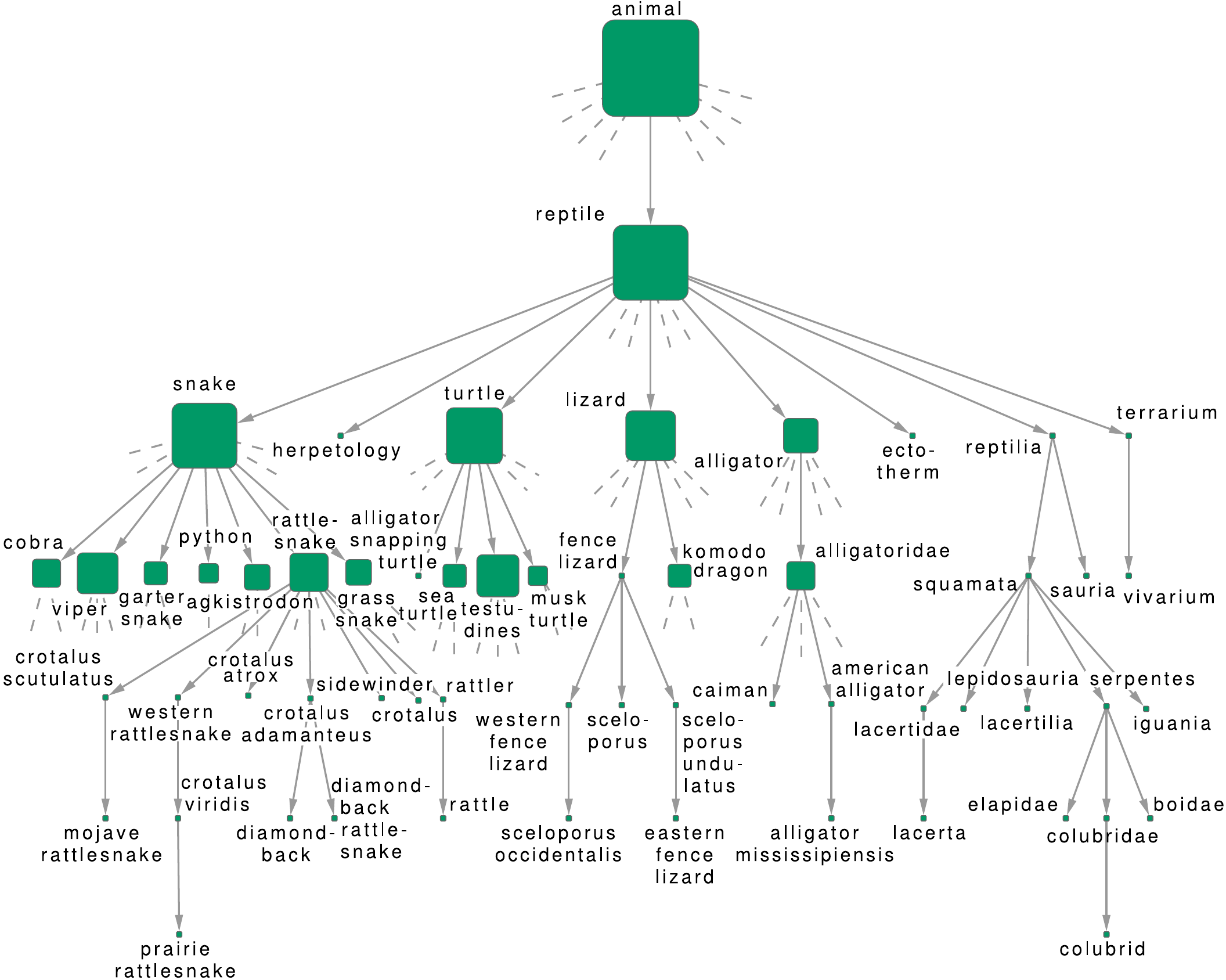}
\end{center}
\begin{flushleft}
\caption{
{\bf Subgraph from the hierarchy between Flickr tags.} By running our algorithm B on a filtered sample from Flickr, we obtained a hierarchy between the tags appearing on the photos in the sample. Since the total number of tags in our data reached 25,441, here we show only a smaller subgraph from the result, corresponding to a part of the tags categorized under ``reptile''. Stubs correspond to further direct descendants not shown in the figure, and the size of the nodes indicate the total number of descendants on a logarithmic scale, (e.g., ``prairie rattlesnake'' has none, while ``snake'' has altogether 110.). 
\label{fig:Flickr}
}
\end{flushleft}
\end{figure}

An example is given in  Fig.\ref{fig:Flickr}., showing a few descendants of the tag ``reptile'' in our reconstruction. Most important direct descendants are ``snake'', ``lizard'', ``alligator'' and ``turtle''. The tags under these main categories seem to be correctly classified, e.g., ``alligator snapping turtle'' is under ``turtle'', (instead of the also related  ``alligator''). Interestingly, Latin names (binomial names) from the taxonomy of ``reptilia'' form a further individual branch under ``reptile'', however, occasionally we can also see binomial names directly connected to the corresponding English name of the given species. More examples from our result on the Flickr data are given in Sect.S3.1 in the Supporting Information, which taken together with Fig.\ref{fig:Flickr} give an overall impression of a meaningful hierarchy, following the ``common sense'' by and large. (Furthermore, similar samples from the hierarchies extracted by the other methods are also given in Sect.S3.2 in the Supporting Information.)

\subsubsection*{Hierarchy of IMDb tags}
\begin{figure}[!ht]
\begin{center}
\includegraphics[width=0.9\textwidth]{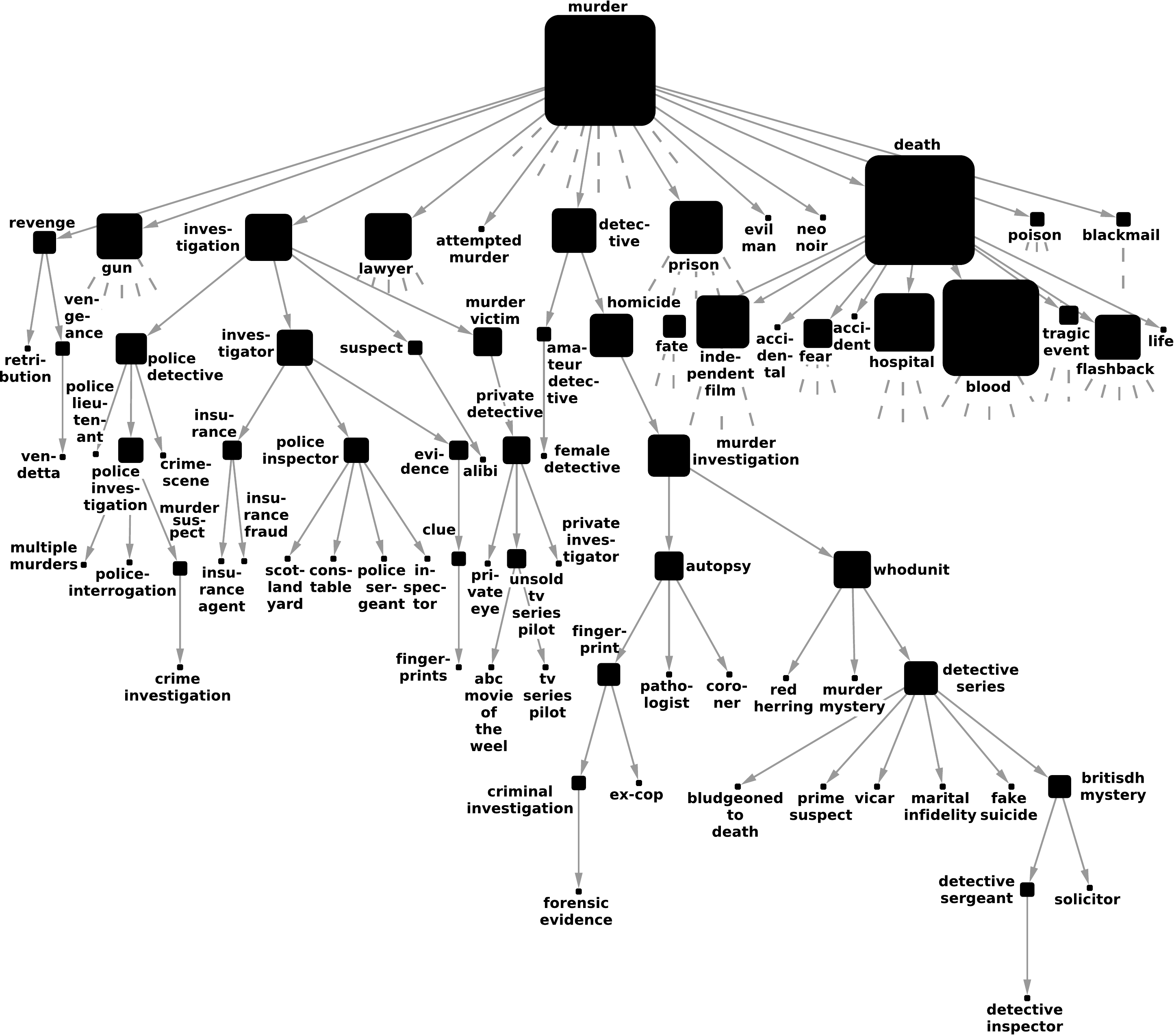}
\end{center}
\caption{{\bf Subgraph from the hierarchy between IMDb tags.} The results were obtained by running Algorithm B on a filtered sample of films from IMDb, tagged by keywords describing the content of the movies. Here we show only a smaller subgraph between the descendants of ``murder'', where stubs correspond to further direct descendants not shown in the figure, and the size of the nodes indicate the total number of descendants on a logarithmic scale.
\label{fig:IMDb}
}
\end{figure}

Another widely known online database is given by the IMDb, providing detailed information related to films, television programs and video games. One of the features relevant from the point of view of our research is that keywords related to the genre, content, subject, scenes, and basically any relevant feature of the movies are also available. These can be treated similarly to the Flickr tags, i.e., they are corresponding to English words, which are not organized into a hierarchy. In Fig.\ref{fig:IMDb}. we show results obtained by running Algorithm B on a relatively large, filtered sample of tagged movies. (The details of the construction of the data set are given in Methods). Similarly to the Flickr data, we display a smaller part of the branch under the tag ``murder'' in the extracted hierarchy. Most important direct descendants are corresponding to ``death'', ``prison'' and ``investigation'', with ``blood'', ``suspect'' and ``police detective'' appearing on lower levels of the hierarchy. Although the tags appearing in the different sub-branches are all related to their parents, the quality of the Flickr hierarchy seemed a bit better. This may be due to the fact that keywords can pertain to any part of the movies, and hence, the tags on a single movie can already be very diverse, providing a more difficult input data set for tag hierarchy extraction. Nevertheless, this result reassures our statement related to the Flickr data, namely that the hierarchies obtained from our algorithm have a meaningful overall impression. (Similar samples from the hierarchies obtained with the other methods are shown in Sect.S3.2 in the Supporting Information.)

\subsection*{Synthetic benchmark based on random walks}
\subsubsection*{Defining the benchmark system}
Providing adjustable benchmarks is very important when testing and comparing algorithms. The basic idea of a benchmark in general is given by a system, where the ground truth about the object of search is also known. However, for most real systems this sort of information is not available, therefore, synthetic benchmarks are constructed. E.g., community finding is one of the very intensively studied area of complex network research, with an enormous number of different community finding algorithms available \cite{Fortunato_report}. Since the ground truth communities are known only for a couple of small networks, the testing is usually carried out on the LFR benchmark \cite{LFR_benchmark}, which is a purely synthetic, computer generated benchmark: the communities are pre-defined, and the links building up the network are generated at random, with linking probabilities taking into account the community structure. The drawback of such synthetic test data is its artificial nature, however, the benefit on the other side is the freedom of the choice of the parameters, enabling the variance of the test conditions on a much larger scale compared to real systems.

Here we propose a similar synthetic benchmark system for testing tag hierarchy extraction algorithms. The basic idea is to start from a given pre-defined hierarchy, (the ``exact'' hierarchy), and generate collections of tags at random, (corresponding to tagged objects in a real system), based on this hierarchy. The tag hierarchy extraction methods to be tested can be run on these sets of tags, and the obtained hierarchies, (the "reconstructed" hierarchies), can be compared to the exact hierarchy used when generating the synthetic data. When drawing an analogy between this system and the LFR benchmark, our pre-defined hierarchy is corresponding to the pre-defined community structure in the LFR benchmark, while the generated collections of tags are corresponding to the random networks generated according to the communities.

\begin{figure}[!ht]
\begin{center}
\includegraphics[width=0.5\textwidth]{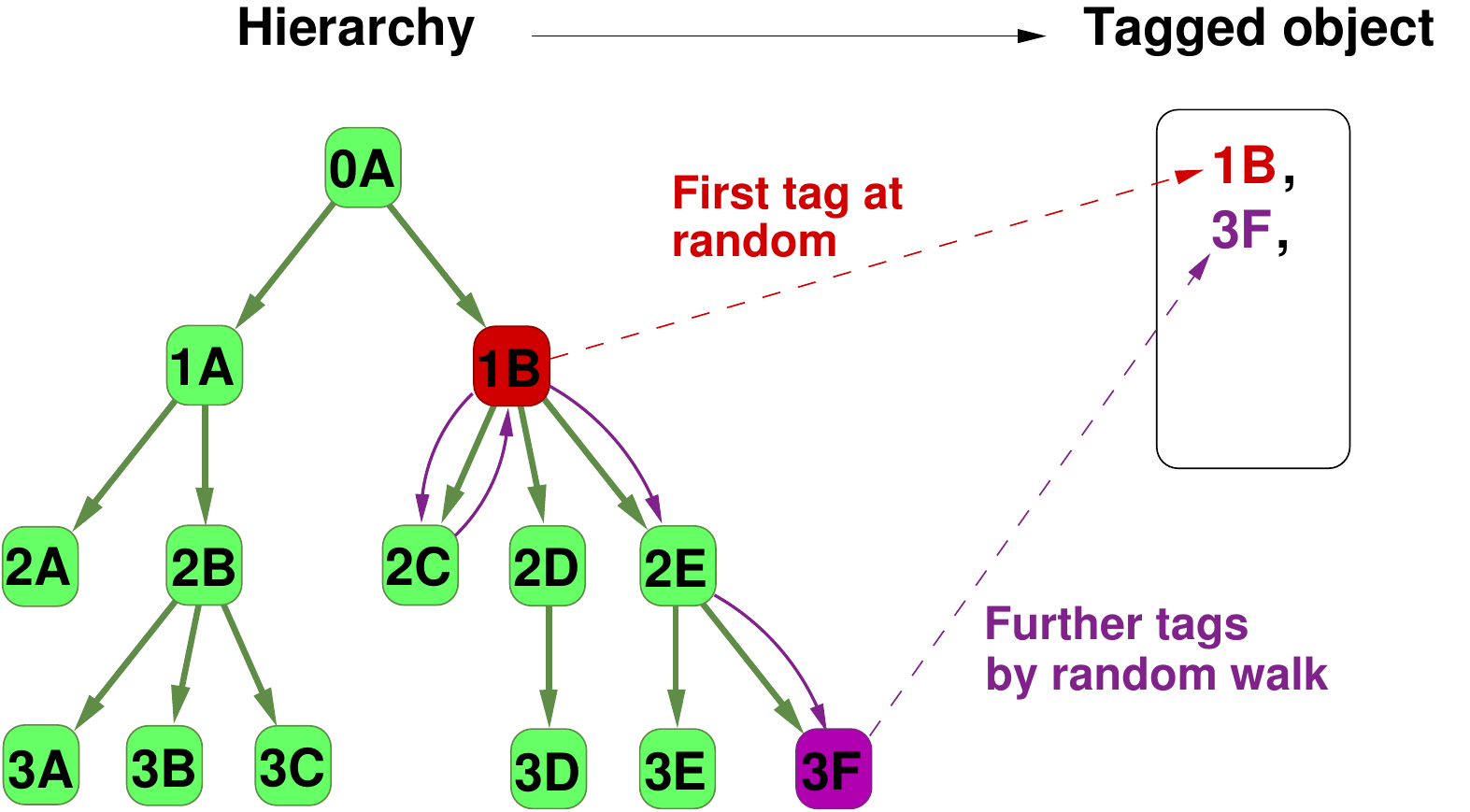}
\end{center}
\caption{
{\bf Generating tags on virtual objects by random walks on the hierarchy.} The objects in this approach are represented simply by collections of tags. For a given collection, the first tag is picked at random, (illustrated in red), while the rest of the tags are obtained by implementing a short undirected random walk on the DAG, starting from the first tag, (illustrated in purple).
\label{fig:random_walk}
}
\end{figure}

To make the above idea of a synthetic tagging system work in practice, we have to specify the method for generating the random collections of tags based on the given pre-defined hierarchy. In general, the basic idea is that tags more closely related to each other according to the hierarchy should appear together with a larger probability compared to unrelated tags. To implement this, we have chosen a random walk approach as suggested in \cite{our_ontology}. The first tag in each collection is chosen at random. For the rest of the tags in the same collection, with probability $p_{\rm RW}$ we start a short undirected random walk on the hierarchy starting from the first tag, and choose the endpoint of the random walk, or with probability $1-p_{\rm RW}$ we again choose at random. An illustration of this process is given in Fig.\ref{fig:random_walk}, (a brief pseudo-code of the data generation algorithm is given in Algorithm S4. in the Supporting Information). The parameters of the benchmark are the following: the pre-defined hierarchy between the tags, the frequency of the tags when choosing at random, the probability $p_{\rm RW}$ for generating the second and further tags by random walk, the length of the random walks, the number of objects and finally, the distribution of the number of tags per object. Although this is a long list of parameters, the quality of the reconstructed hierarchy is not equally sensitive to all of them. E.g., according to our experiments change in the topology of the exact hierarchy, or in the length of the random walk have only a minor effect, while the distribution of the tag frequencies seems to play a very important role.

\subsubsection*{Results on synthetic data}

In Table \ref{table:comp_gen_1}. we show the tag hierarchy extraction results obtained on synthetic data generated by using our random walk based benchmark system. In the data generation process the exact hierarchy was set to a binary tree of 1,023 tags, with tag frequencies decreasing linearly as a function of the depth in the hierarchy. We generated  an average number of 3 co-occurring tags on altogether 2,000,000 hypothetical objects, with random walk probability of $p_{\rm RW}=0.5$ and random walk lengths chosen from a uniform distribution between 1 and 3. We ran the same algorithms on the obtained data as in case of the protein data set, and used the same measures for evaluating the quality of the results. According to Table \ref{table:comp_gen_1}., the majority of the algorithms perform very well, e.g., algorithm B and the algorithm by P.~Heymann \& H.~Garcia-Molina are producing almost perfect reconstructions, thus, this example is an ``easy'' data set. Interestingly, the results of the algorithm by Schmitz were very poor on this input. Nevertheless, this method is still competitive with the others, e.g., it showed a quite good performance on the protein data set. However, the study of why does this algorithm behave completely different from the others on our benchmark is out of the scope of the present work.

\begin{table}[!ht]
\caption{
\bf{Quality measures of the reconstructed hierarchies for the ``easy'' synthetic data set.}}
\begin{tabular}{|l|c|c|c|c|c|c|c|}
\hline
    & $r_{\rm E}$ & $r_{\rm A}$ & $r_{\rm I}$ & $r_{\rm U}$ & $r_{\rm M}$ & $I_{\rm e,r}$ & $I_{\rm lin}$
 \\ \hline  
algorithm A & 67\%   &100\% & 0\% & 0\%& 0\% & 91\% & 99\% \\
\hline
algorithm B &  100\% & 100\% & 0\% & 0\% & 0\% & 100\% & 100\% \\
\hline 
P.~Heymann \& H.~Garcia-Molina & 99\% & 100\% & 0\% & 0\%&  0\%& 93\% & 99\% \\
\hline
P.~Schmitz &  0\% & 0\% & 0\% & 0\% & 100\% & 0\% & 0\% \\
\hline
\end{tabular}
\begin{flushleft}
When generating the data set, the frequency of the initial tags was decreasing linearly as a function of the level depth in the exact hierarchy. We show the same quality measures as in Table \ref{table:results}.: the ratio of exactly matching links, $r_{\rm E}$, the  ratio of acceptable links, $r_{\rm A}$, the ratio of inverted links, $r_{\rm I}$, the ratio of unrelated links, $r_{\rm U}$, the ratio of missing links, $r_{\rm M}$, the normalized mutual information between the exact- and the reconstructed hierarchies, $I_{\rm e,r}$, and the linearized mutual information, $I_{\rm lin}$. The different rows correspond to results obtained from algorithm A, (1$^{\rm st}$ row), algorithm B, (2$^{\rm nd}$ row), the method by  P.~Heymann \& H.~Garcia-Molina (3$^{\rm d}$ row), and the algorithm by P.~Schmitz (4$^{\rm th}$ row).
\end{flushleft}
\label{table:comp_gen_1}
\end{table}

The ``easy'' synthetic data discussed above can be turned into a ``hard'' one by changing the frequency distribution of the tags. In Table \ref{table:comp_gen_2}. we show the results obtained when the tag frequencies were independent of the level depth in the hierarchy, and had a power-law distribution, with the other parameters of the benchmark left unchanged. According to the studied quality measures, the performance of the involved methods drops down drastically compared to Table \ref{table:comp_gen_1}. However, algorithm B provides an exception in this case, achieving pretty good results even for this ``hard'' test data. E.g., the NMI value is still $I_{\rm e,r}=0.83$ for our algorithm, while for e.g., the algorithm by P. Heymann \& H. Garcia-Molina it is reduced to $I_{\rm e,r}=0.29$. Moreover, the fraction of exactly matching links is almost 90\% for algorithm B, while it is below 50\% for the algorithm by P. Heymann \& H. Garcia-Molina. This shows that algorithm B can have a significantly better performance compared to other algorithms, as the quality of its output is less dependent on the correlation between tag frequencies and level depth in the hierarchy. Another interesting effect in Table \ref{table:comp_gen_1}. is that the results for the algorithm by Schmitz are slightly better compared to the ``easy'' data set. As we mentioned earlier, studies of the reasons for the outlying behavior of this algorithm on our benchmark compared to the other methods is left for future work.

\begin{table}[!ht]
\caption{
\bf{Quality measures of the reconstructed hierarchies for the ``hard'' synthetic data set.}}
\begin{tabular}{|l|c|c|c|c|c|c|c|}
\hline
    & $r_{\rm E}$ & $r_{\rm A}$ & $r_{\rm I}$ & $r_{\rm U}$ & $r_{\rm M}$ & $I_{\rm e,r}$ & $I_{\rm lin}$
 \\ \hline  
algorithm A & 31\%   &35\% & 18\% & 47\%& 0\% & 18\% & 66\% \\
\hline
algorithm B &  89\% & 91\% & 6\% & 3\% & 0\% & 83\% & 97\% \\
\hline 
P.~Heymann \& H.~Garcia-Molina & 48\% & 54\% & 29\% & 17\%&  0\%& 29\% & 76\% \\
\hline
P.~Schmitz &  1\% & 2\% & 1\% & 3\% & 94\% & 1\% & 5\% \\
\hline
\end{tabular}
\begin{flushleft}
In this case the frequency of the initial tags was independent of their position in the exact hierarchy during the benchmark generation, and the frequency distribution followed a power-law. This change compared to the data set used in Table \ref{table:comp_gen_1}. results in significant decrease in the quality measures for most of the involved methods, as shown by   the  ratio of acceptable links, $r_{\rm A}$, the ratio of inverted links, $r_{\rm I}$, the ratio of unrelated links, $r_{\rm U}$, the ratio of missing links, $r_{\rm M}$, the normalized mutual information between the exact- and the reconstructed hierarchies, $I_{\rm e,r}$, and the linearized mutual information, $I_{\rm lin}$. The different rows correspond to results obtained from algorithm A, (1$^{\rm st}$ row), algorithm B, (2$^{\rm nd}$ row), the method by  P.~Heymann \& H.~Garcia-Molina (3$^{\rm d}$ row), and the algorithm by P.~Schmitz (4$^{\rm th}$ row).
\end{flushleft}
\label{table:comp_gen_2}
\end{table}

 The effects of the modifications in the other parameters of the benchmark are discussed in Sect.S4.2-S4.3 in the Supporting Information. Nevertheless these results already show that the provided framework can serve as versatile test tool for tag hierarchy extraction methods.

\section*{Methods}
\subsection*{$z$-score}
Both algorithms introduced in the paper depend on the $z$-score related to the number of co-occurrences between a pair of tags. If the tags are assigned to the objects completely at random, the distribution of the number of co-occurrences for a given pair of tags $i$ and $j$ follows the hypergeometric distribution: Assuming that tag $i$ and $j$ appear altogether on $Q_i$ and $Q_j$ objects respectively, let us consider the random assignment of tag $i$ among a total number of $Q$ objects. This is equivalent to drawing $Q_i$ times from the objects without replacement, where the ``successful'' draws correspond to objects also having tag $j$, (and the total number of such objects is $Q_j$). Based on this, the probability for observing a given $Q_{ij}$ number of co-occurrences between $i$
 and $j$ is
\begin{equation}
P(Q_{ij}=k)=\frac{\binom{Q_j}{k}\binom{Q-Q_j}{Q_i-k}}{\binom{Q}{Q_i}},
\end{equation}
with the expected number of co-appearances given by
 \begin{equation}
\left< Q_{ij}\right>=\frac{Q_{i}Q_{j}}{Q},
\label{eq:exp_rand}
\end{equation}
and the variance formulated as
\begin{equation}
\sigma^2(Q_{ij})=\frac{Q_iQ_j}{Q}\frac{Q-Q_i}{Q}\frac{Q-Q_j}{Q-1}.
\label{eq:var}
\end{equation}
The $z$-score is defined as the difference between the observed number of co-occurrences in the data, $Q_{ij}$, and the expected number of co-occurrences at random as given in (\ref{eq:exp_rand}), scaled by the standard deviation according to (\ref{eq:var}), 
\begin{equation}
z_{ij}=\frac{Q_{ij}-\left< Q_{ij}\right>}{\sigma(Q_{ij})}.
\end{equation} 

\subsection*{Normalized mutual information}
For discrete variables $x_i$ and $y_j$ with a joint probability distribution given by $P(x_i,y_j)$, the mutual information is defined as 
\begin{equation}
I(x,y)\equiv \sum_{i}\sum_{j} p(x_i,y_j)\ln \left(\frac{p(x_i,y_j)}{p(x_i)p(y_j)}\right), \label{eq:mutinfo}
\end{equation} 
where $p(x_i)$ and $p(y_j)$ denote the (marginal) probability distributions of $x_i$ and $y_j$ respectively. If the two variables are independent, $p(x_i,y_j)=p(x_i)p(y_j)$, thus, $I(x,y)$ becomes $0$. The above quantity is very closely related to the entropy of the random variables, 
\begin{equation}
I(x,y)=H(x)+H(y)-H(x,y), \label{eq:mutinfo_entropy}
\end{equation}
where $H(x)=-\sum_ip(x_i)\ln p(x_i)$ and $H(y)=-\sum_j p(y_j)\ln p(y_j)$ correspond to the entropy of $x$ and $y$, while $H(x,y)=-\sum_{ij}p(x_i,y_j)\ln p(x_i,y_j)$ denotes the joint entropy. Based on (\ref{eq:mutinfo_entropy}), the NMI can be defined as
\begin{equation}
I_{\rm norm}(x,y)\equiv \frac{2 I(x,y)}{H(x)+H(y)}. \label{eq:NMI_alap}
\end{equation}
This way the NMI is 1 if and only $x$ and $y$ are identical, and 0 if they are independent.

\subsection*{Data}
\subsubsection*{Protein data}
Both the exact DAG describing the hierarchy between protein functions and the corresponding input data set given by proteins tagged with known function annotations were taken from the Gene Ontology \cite{GO}. The hierarchy of protein function is composed of three separate DAGs, corresponding to ``molecular function'', ``biological process'' and ``cellular component''. We concentrated on molecular functions, where the complete DAG has altogether 6,469 tags. However, a considerable part of these annotations are rather rare, thus, reconstructing the complete hierarchy would be a very hard task due to the lack of information. Therefore, we took a smaller subgraph, namely the branch starting from ``catalytic activity'', counting 4,181 tags, most of which are relatively more frequent. 

For the data set of proteins, tagged with their known molecular function annotations,  we took the monthly (quality controlled) release as in 2012.08.01. For simplicity, we neglected proteins lacking any tags appearing in the exact hierarchy, and deleted all annotations which are not descendants of ``catalytic activity''. The resulting smaller data set contained 5,913,610 proteins, each having on average 3.7 tags. This data set, (together with the corresponding exact DAG) is available at (http://hiertags.elte.hu).

\subsubsection*{Flickr data}
Flickr provides the possibility for searching photos by tags, thus, as a first step we downloaded photos resulting from search queries over a list of 68,812 English nouns, yielding altogether 2,565,501 photos, (the same photo can appear multiple times as a result for the different queries). At this stage we stored all the tags of the photos and the anonymous user id of the photo owners as well. Next, the set of tags on the photos had to be cleaned: only English nouns were accepted, and in case of parts of a compound word appeared beside the compound word on the same photo, the smaller parts were deleted, leaving only the complete compound word. Since our algorithms rely on the weighted network of co-appearances, we applied a further filtering: a link was accepted only if the corresponding tags co-appeared on photos belonging to at least 10 different users. The resulting tag co-appearance network had 25,441 nodes, encoding information originating from 1,519,030  photos. We made the list of weighted links between the tags available at (http://hiertags.elte.hu).

\subsubsection*{IMDb data}
We have downloaded the data from the IMDb Web site\cite{imdbwww}, and used the ``keywords.list.gz'' data file, listing the keywords associated with the different movies. The goal of the keywords is helping the users in searching amongst the movies, and keywords can pertain to any part, scene, subject, gender, etc. of the movie. Although keywords can be given only by registered users, there is no restriction what so ever for registering, and the submitted information is processed by the "Database Content Team" of the IMDb site. The version of the original data we are used here contained 487,356 movie titles and 136,204 different keywords. However, to improve the quality of the data set, we restricted our studies to keywords appearing on at least a 100 different movies, leaving 336,223 movies and 6,358 different keywords in the data set. This cleaned version is available at (http://hiertags.elte.hu).

\section*{Discussion}
We introduced a detailed framework for tag hierarchy extraction in tagging systems. First, we have defined quality measures for hierarchy extraction methods based on comparing the obtained results to a pre-defined exact hierarchy. A part of these quantities were simply given by fractions of links fulfilling some criteria, (e.g., exactly matching, inverted, etc.). However we also defined the NMI between the exact- and the reconstructed hierarchies, providing a quality measure which is sensitive also to the position of the non-matching links in the hierarchy. This was illustrated by our experiments comparing a hierarchy to its randomized counterpart, where the NMI showed a significantly faster decay when the rewiring was started at the top of the hierarchy, compared to the opposite case of starting from the leafs. 

Furthermore, we developed a synthetic, computer generated benchmark system for tag hierarchy extraction. This tool provides versatile possibilities for testing hierarchy extraction algorithms under controlled conditions. The basic idea of our benchmark is generating collections of tags associated to virtual objects based on a pre-defined hierarchy between the tags. By running a tag hierarchy extraction algorithm on the generated synthetic data, the obtained result can be compared to the pre-defined exact hierarchy used in the data generation process. According to our experiments on the benchmark,  by changing the parameters during the generation of the synthetic data, we can enhance or decrease the difficulty of the tag hierarchy reconstruction.

 In addition, we developed two novel tag hierarchy extraction algorithms based on the network approach, and tested them both on real systems and computer generated benchmarks. In case of the tagged protein data the similarity between the obtained protein function hierarchy and the hierarchy given by the Gene Ontology was very encouraging, and the hierarchy between the English words obtained for the Flickr and IMDb data sets seemed also quite meaningful. The computer generated benchmark system we have set up provides further possibilities for testing tag hierarchy extraction algorithms in general. By changing the parameters during the input generation we can enhance or decrease the difficulty of the tag hierarchy reconstruction. 
 
Our methods were compared to current state of the art tag hierarchy extraction algorithms by P.~Heymann \& H.~Garcia-Molina and by P.~Schmitz. Interestingly, the rank of the algorithms according to the introduced quality measures was varying from system to system. In case of the protein data set algorithm A was slightly ahead of the others, while the rest of the methods achieved more or less the same quality. In turn, for the easy synthetic test data, algorithm B and the algorithm by P.~Heymann \& H.~Garcia-Molina reached almost perfect reconstruction, with algorithm A left slightly behind, and the algorithm by P. Schmitz achieving very poor marks. However, when changing to the hard synthetic test data, a large difference was observed between the quality of the obtained results, as algorithm B significantly outperformed all other methods.

The different ranking of the algorithms for the included examples indicates that tag hierarchy extraction is a non-trivial problem where a system can be challenging for one given approach and easy for another method and vice versa. Nevertheless the results obtained indicate that tag hierarchy extraction is a very promising direction for further research with a great potential for practical applications.


\input{Hiertags_arxiv1.bbl}
\newpage

\input{Hiertags_SI_resub_01}

\end{document}

%% file: Hiertags_SI_resub_01.tex
\makeatletter

\makeatother
\renewcommand{\thefigure}{S\arabic{figure}}
\renewcommand{\thetable}{S\arabic{table}}
\renewcommand{\theequation}{S\arabic{equation}}
\renewcommand{\thealgorithm}{S\arabic{algorithm}}
\renewcommand{\thesection}{S\arabic{section}}

\begin{center}
{\LARGE
\textbf{Extracting tag-hierarchies}}\\
{\Large \textbf{Supporting Information}
}

\vspace{1.5cm}

\end{center}

\section{Algorithms}

\subsection{Algorithm A}
\subsubsection{Complexity}
The pseudo code of the algorithm would be to long to be displayed in a single page, thus, we divided it into two parts. The first part, corresponding to the preparation of the weighted network between the tags and the building of local hierarchies is given in Algorithm \ref{algA_first}.
\begin{algorithm}[!ht]
\caption{Algorithm A, 1$^{\rm st}$ part: building local hierarchies.}
\label{algA_first}
\begin{algorithmic}[1]
\ForAll{objects: object1} 
        \ForAll{tags appearing on object1: tag1} 
               \ForAll{tags appearing on object1: tag2} 
               coappearances(tag1,tag2)+=1
               \EndFor
        \EndFor
\EndFor
\ForAll{tags: tag1}
        \State max= maximal coappearances(tag1,tag2)
	\ForAll{tags: tag2}
		\If{coappearances(tag1, tag2) $>=$ $\omega$ * max}
                        \State calc zscore(tag1,tag2)
			\State strongpartners(tag1, tag2) = zscore(tag1, tag2)
		\EndIf
	\EndFor
\EndFor
\ForAll{tags: tag1}
	\State parent = undef
	\ForAll{strongpartners of tag1 sorted to descending order: tag2}
		\If{parent = undef and not exists strongpartners(tag2, tag1)}
			\State parent = tag2
		\EndIf
	\EndFor
\EndFor
\end{algorithmic}
\end{algorithm}
By assuming that the number of tags on one object is $\mathcal{O}(1)$, the number of operations needed for generating the weighted co-occurrence network between the tags can be given by the number of objects, $Q$, as $\mathcal{O}(Q)$. According to our experience, the resulting co-occurrence network between the tags is usually sparse, thus, the number of links in the network between the tags, $M$, and the number of tags, $N$, are similar in magnitude, $\mathcal{O}(M)=\mathcal{O}(N)$, and the average number of links of the tags is $\mathcal{O}(1)$. According to that, the individual thresholding of the network based on the strongest link on each tag also needs $\mathcal{O}(N \log N)$ operations. Similarly, the calculation of the $z$-score and choosing the in-neighbor with the highest value as a parent need also $\mathcal{O}(M \log M)$ operations.

 In the next phase, the smaller isolated subgraphs under the local roots have to be assembled into a single hierarchy, as shown in Algorithm \ref{algA_sec}.
\begin{algorithm}[!ht]
\caption{Algorithm A, 2$^{\rm nd}$ part: assembly into a global hierarchy.}
\label{algA_sec}
\begin{algorithmic}[1]
\If{there are more components}
	\ForAll{roots: root}
		\State h(root) = entropy of root
		\ForAll{tags in the component of root: tag1}
			\State component(tag1) = root
		\EndFor
	\EndFor
	\State global$\_$root = root with highest entropy
	\ForAll{roots except the global: root}
		\State suggested$\_$parent(root) = undef
		\ForAll{coappearing tags sorted to descending coappearances: tag2}
			\If{suggested$\_$parent(root) = undef and component(tag2) is not root}
				\State suggested$\_$parent(root) = tag2
			\EndIf
		\EndFor
	\EndFor
	\ForAll{roots appearing in suggested$\_$parent: root}
		\State tag1 = root
		\State empty visited
		\While{does not exists visited(tag1) and exists suggested$\_$parent(tag1)}
			\State tag1 = component(suggested$\_$parent(tag1))
			\State visited(tag1) = 1
		\EndWhile
		\If{exists visited(tag1)}
			\ForAll{roots in visited: root2}
				looped(root2) = 1
				delete suggested$\_$parent(root2)
			\EndFor
		\EndIf
	\EndFor
	\ForAll{roots in looped, sorted to descending order of h: root}
		\ForAll{tags coappearing with root: tag1}
			\If{not exists suggested$\_$parent(root)}
				\State check whether tag1 is below root
				\If{tag1 is not below root}
					suggested$\_$parent(root) = tag1
				\EndIf
			\EndIf
		\EndFor
		\If{not exists suggested$\_$parent(root)}
			suggested$\_$parent(root) = global$\_$root
		\EndIf
	\EndFor
\EndIf

\end{algorithmic}
\end{algorithm}
 Choosing the global root of the hierarchy needs $\mathcal{O}(N)$ operations, and similarly, choosing the parent of a local root also needs at most $\mathcal{O}(N)$ operations. During this process we need to detect (and correct) possible newly created loops, requiring at most $\mathcal{O}(N)$ operations. Based on the above, the resulting overall complexity of algorithm A is $\mathcal{O}(Q) + \mathcal{O}(M \log M)= \mathcal{O}(Q)+\mathcal{O}(N \log N)$, where we assumed that the co-occurrence network between the tags is sparse, i.e., $\mathcal{O}(M)=\mathcal{O}(N)$.

\subsubsection{Optimizing the parameter $\omega$}
\label{sect:optim_omega}
\begin{figure}[!ht]
\begin{center}
\includegraphics[width=0.8\textwidth]{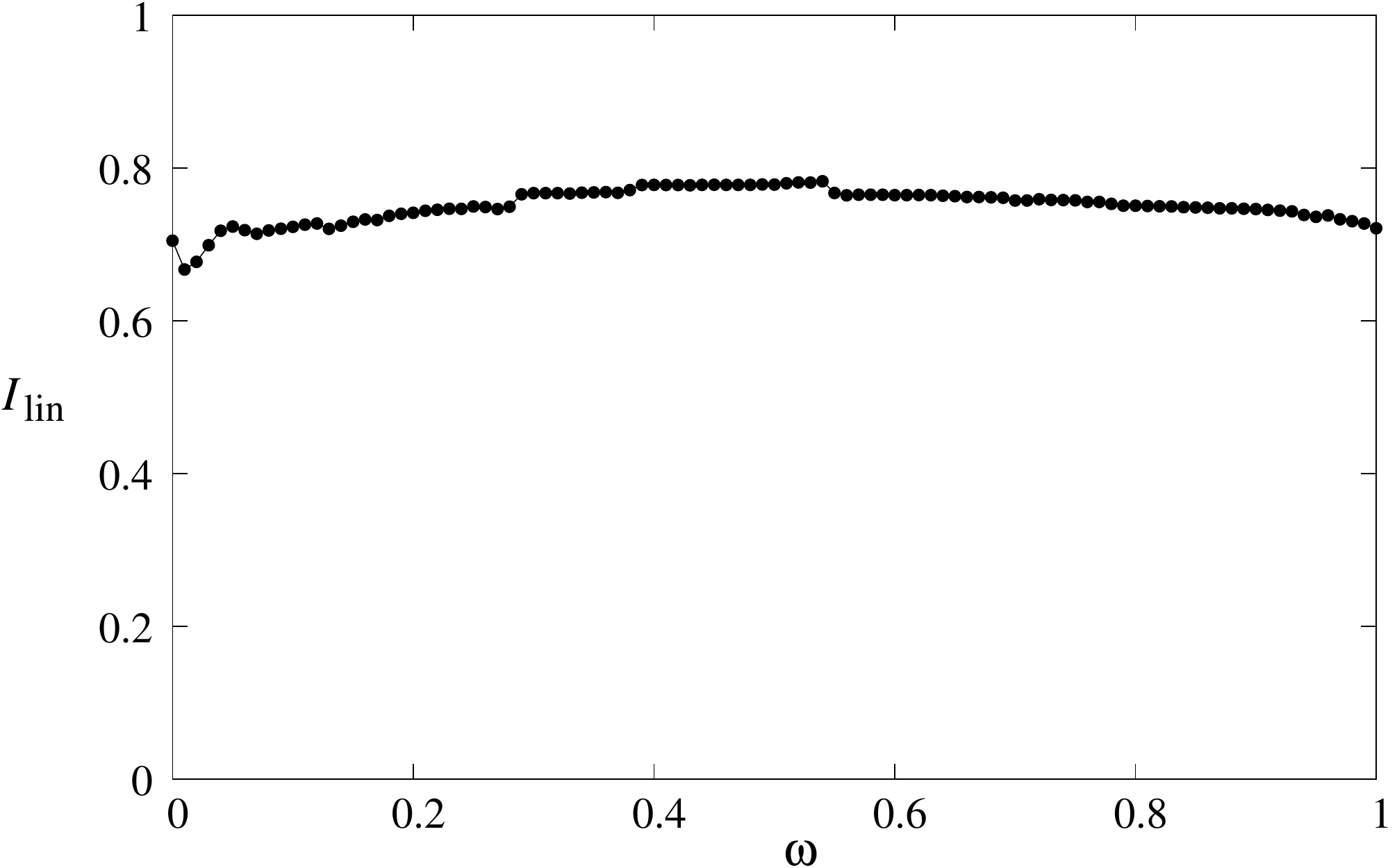}
\end{center}
\begin{flushleft}
\caption{
{\bf Optimizing the parameter $\omega$.} 
We show the LMI as a function of $\omega$ for the protein function data set.
\label{fig:opt_omega}
}
\end{flushleft}
\end{figure}

The parameter $\omega \in[0,1]$ in algorithm A is corresponding to the local weight threshold used for throwing away weak connections in the tag co-occurrence network. In order to find the optimal value for $\omega$, we measured the LMI between the reconstructed hierarchy and the exact hierarchy as a function of $\omega$ in case of the protein function data set. The results of this experiment are shown in Fig.\ref{fig:opt_omega}. Although $I_{\rm lin}$ is showing only minor changes over the whole range of possible $\omega$ values, a maximal plateau can still be observed between $\omega=0.3$ and $\omega=0.55$. Based on this, throughout the experiments detailed in our paper, we used algorithm A with $\omega=0.4$.

\subsection{Algorithm B} 
\subsubsection{Complexity}
\begin{algorithm}[!ht]
\caption{Algorithm B}
\label{algB}
\begin{algorithmic}[1]
\ForAll{tags: tag1}
	\ForAll{tags: tag2}
		\State calc zscore(tag1, tag2)
	\EndFor
\EndFor
\ForAll{tags: tag1}
	\ForAll{tags: tag2}
		\If{zscore(tag1, tag2) $>$ threshold$\_$B or coappearances(tag1, tag2) $>=$ 0.5 * objects(tag1) or coappearances(tag1, tag2) $>=$ 0.5 * objects(tag2)}
			\State M(tag1, tag2) = coappearances(tag1, tag2)
			\State strength(tag1) += coappearances(tag1, tag2)
		\EndIf
	\EndFor
\EndFor
\ForAll{tags: tag1}
	\State centrality(tag1) = strength(tag1)
\EndFor
\For{i=1, i$<=$100}
	\State sum = 0
	\ForAll{tags: tag1}
		\ForAll{tags: tag2}
			\State temp$\_$centrality(tag1) = M(tag1, tag2) * centrality(tag2)
		\EndFor
		\State sum += temp$\_$centrality(tag1)
	\EndFor
	\ForAll{tags: tag1}
		centrality(tag1) = temp$\_$centrality(tag1) / sum
	\EndFor
\EndFor
\For{tags sorted to ascending centralities: tag1}
	\State empty score;
	\For{coappearing partners of tag1: tag2}
		\State score(tag2) = zscore(tag1, tag2)
	\EndFor
	\For{descendants of tag1: desc}
		\For{coappearing partners of desc: tag2}
			\If{tag2 coappears with tag1 and centrality(tag2) > centrality(tag1) and (zscore(tag1, tag2) $>$ threshold$\_$B or coappearances(tag1, tag2) $>=$ 0.5 * objects(tag1)) and (zscore(desc, tag2) $>$ threshold$\_$B or coappearances(desc, tag2) $>=$ 0.5 * objects(desc))}
				\State score(tag2) += zscore(desc, tag2)
			\EndIf
		\EndFor
	\EndFor
	\If{score is not empty}
		\State parent(tag1) = highest scoring tag
	\EndIf
\EndFor
\end{algorithmic}
\end{algorithm}
The pseudo code for the algorithm is given in Algorithm \ref{algB}.
The preparation of the tag co-occurrence network is the same as in case of algorithm A, with a complexity of $\mathcal{O}(Q)$, and similarly, the calculation of the $z$-score needs $\mathcal{O}(M)$ operations. To evaluate the eigenvector centrality, we simply use the power iteration method on the filtered co-appearance matrix, (see the pseudo code), which needs $\mathcal{O}(N)$ operations, for the typical case of a sparse matrix. The hierarchy is assembled bottom up, and the calculation of the scores for the possible parents of a given tag requires $\mathcal{O}(N\cdot\log N)$ operations, assuming that the structure of the complete DAG is similar to a tree with a constant branching number. (In case it is chain-like, this is modified to $\mathcal{O}(N^2)$, whereas for a star-like topology, it is only $\mathcal{O}(N)$). The resulting overall complexity of the algorithm is $\mathcal{O}(Q) + \mathcal{O}(N\cdot\log N)$.

\subsubsection{Optimizing the $z$-score threshold}
\begin{figure}[!ht]
\begin{center}
\includegraphics[width=0.8\textwidth]{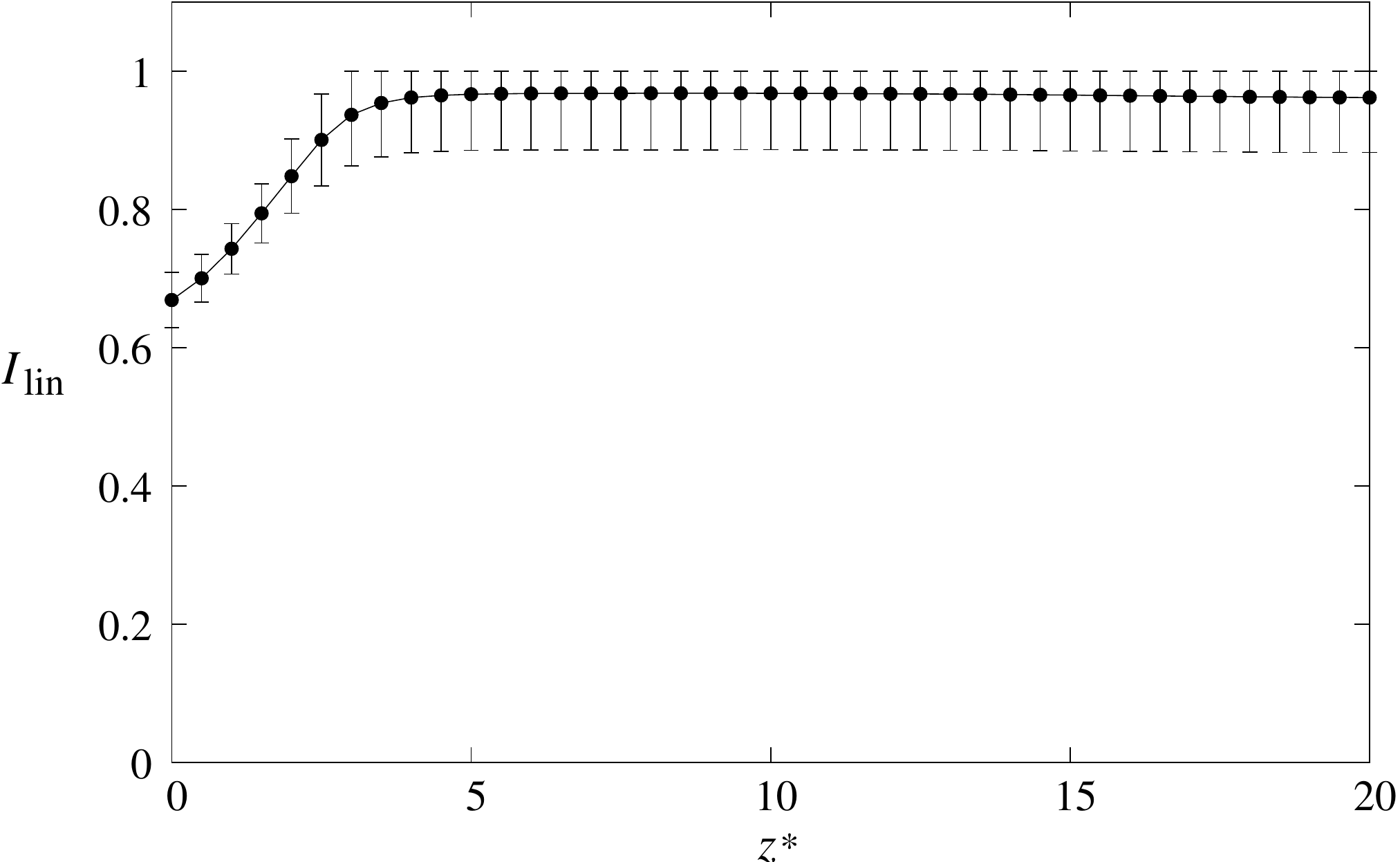}
\end{center}
\begin{flushleft}
\caption{
{\bf Optimizing the $z$-score threshold.} 
We show the LMI as a function of the $z$-score threshold for the ``hard'' synthetic data set.
\label{fig:opt_z}
}
\end{flushleft}
\end{figure}

The $z$-score threshold is an important parameter in algorithm B, which is used for pruning the network of co-occurrences between the tags by throwing away irrelevant connections. In order to optimize this parameter, we have run tests on the ``hard'' synthetic data set, introduced in Section ``Results on synthetic data'' in the main paper. The reason for this choice instead of, e.g., the protein function data set as in Sect.\ref{sect:optim_omega}, is that algorithm B showed best performance on this data set. In Fig.\ref{fig:opt_z}.\ we show the LMI between the reconstructed hierarchy and the exact hierarchy as a function of the $z$-score threshold $z^*$. Although the obtained curve is rather flat in most of the examined region, setting the threshold to $z^*=10$ in general seems as a good choice: below $z^*=5$ the quality drops down, whereas no significant increase can be observed in $I_{\rm lin}$ between $z^*=10$ and $z^*=20$. By choosing $z^*=10$, we ensure good quality, and also avoid throwing away too many connections.

\section{Normalized mutual information}
\subsection{NMI by partitioning of the tags}
\begin{figure}[!ht]
\begin{center}
\includegraphics[width=0.8\textwidth]{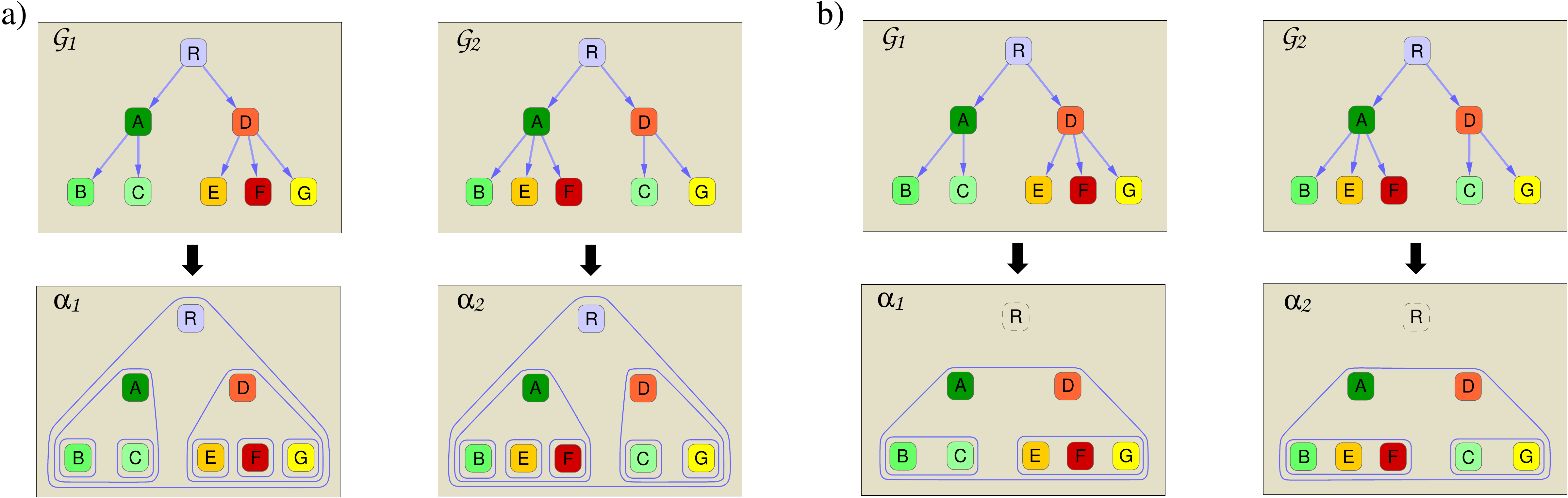}
\end{center}
\begin{flushleft}
\caption{
{\bf Mapping from hierarchies to communities.} a) A simple intuitive mapping from the DAG to a communities of the tags in the DAG is given by nested sets, as shown here for $\mathcal{G}_1$ and $\mathcal{G}_2$, resulting in partitions $\alpha_1$ and $\alpha_2$. b) If we use instead communities given by the union of all descendants from non-leaf tags, (always excluding the given tag itself), the NMI given by (\ref{eq:NMI_com}) becomes equivalent to the NMI defined for hierarchies in Eq.(1) in the main paper.
\label{fig:NMI_com}
}
\end{flushleft}
\end{figure}
As mentioned in the main paper, a very important application of the concept of the NMI is given in community detection, where this measure can be used to quantify the similarity between partitions of the same network into communities by two alternative methods \cite{Danon_mutinfo,Lancichinetti_mutinfo}. The formula providing the NMI between community partitions $\alpha$ and $\beta$ can be given as\begin{equation}
I_{\alpha,\beta}=\frac{-2\sum\limits_{i=1}^{C_{\alpha}}\sum\limits_{j=1}^{C_{\beta}}N_{ij}\ln\left(\frac{N_{ij}N}{N_iN_j}\right)}{\sum\limits_{i=1}^{C_{\alpha}}N_i\ln \left(\frac{N_i}{N}\right)+\sum\limits_{j=1}^{C_{\beta}}N_j\ln \left(\frac{N_j}{N}\right)},
\label{eq:NMI_com}
\end{equation}
where $C_{\alpha}$ and $C_{\beta}$ denote the number of communities in the two partitions, $N_i$ and $N_j$ stand for the number of nodes in communities $i$ and $j$ respectively, with $N_{ij}$ giving the number of common nodes in $i$ and $j$, and finally, $N$ denoting the total number of nodes in the network. This measure can be used e.g., when judging the quality of a community finding method run on a benchmark for which the ground truth communities are known.

Meanwhile, (\ref{eq:NMI_com}) is in complete analogy with our definition of the NMI for a pair of hierarchies, (Eq.(1) in the main paper): if we convert the hierarchies to be compared into community partitions in an appropriate way, the two measures become equivalent. Probably the most natural idea for a mapping from a DAG to communities of the tags in the DAG is turning the original ``order'' hierarchy represented by the DAG into a ``containment'' hierarchy of nested sets, as shown in Fig.\ref{fig:NMI_com}a., (with each set corresponding to the union of tags in a given branch of the DAG). However, by applying (\ref{eq:NMI_com}) to the partitions obtained in this way we obtain different results compared to Eq.(1) in the main paper, and the resulting similarity measure does not approach 0 even for independent random DAGs. The reason for this effect is that leafs in the DAG provide communities consisting of single nodes, and due to the relatively large number of leafs in a general DAG, we always obtain a non vanishing portion of exactly matching communities.

The mapping from a DAG to communities providing results equivalent to our NMI definition is obtained by associating with every tag in the DAG the union of its descendants, excluding the tag itself, (see Fig.\ref{fig:NMI_com}b for illustration). This way the leafs appear only in the communities corresponding to their ancestors, thus, the emergence of a large number of communities with only a single member is avoided.

\subsection{Gene Ontology DAG}
In Fig.1. in the main paper we have examined the behavior of the NMI between a binary tree and its randomized counterpart as a function of the fraction of rewired links. Here we show similar results obtained for the exact hierarchy of our protein data set, obtained from the Gene Ontology \cite{GO}.
\begin{figure}[!ht]
\begin{center}
\includegraphics[width=0.7\textwidth]{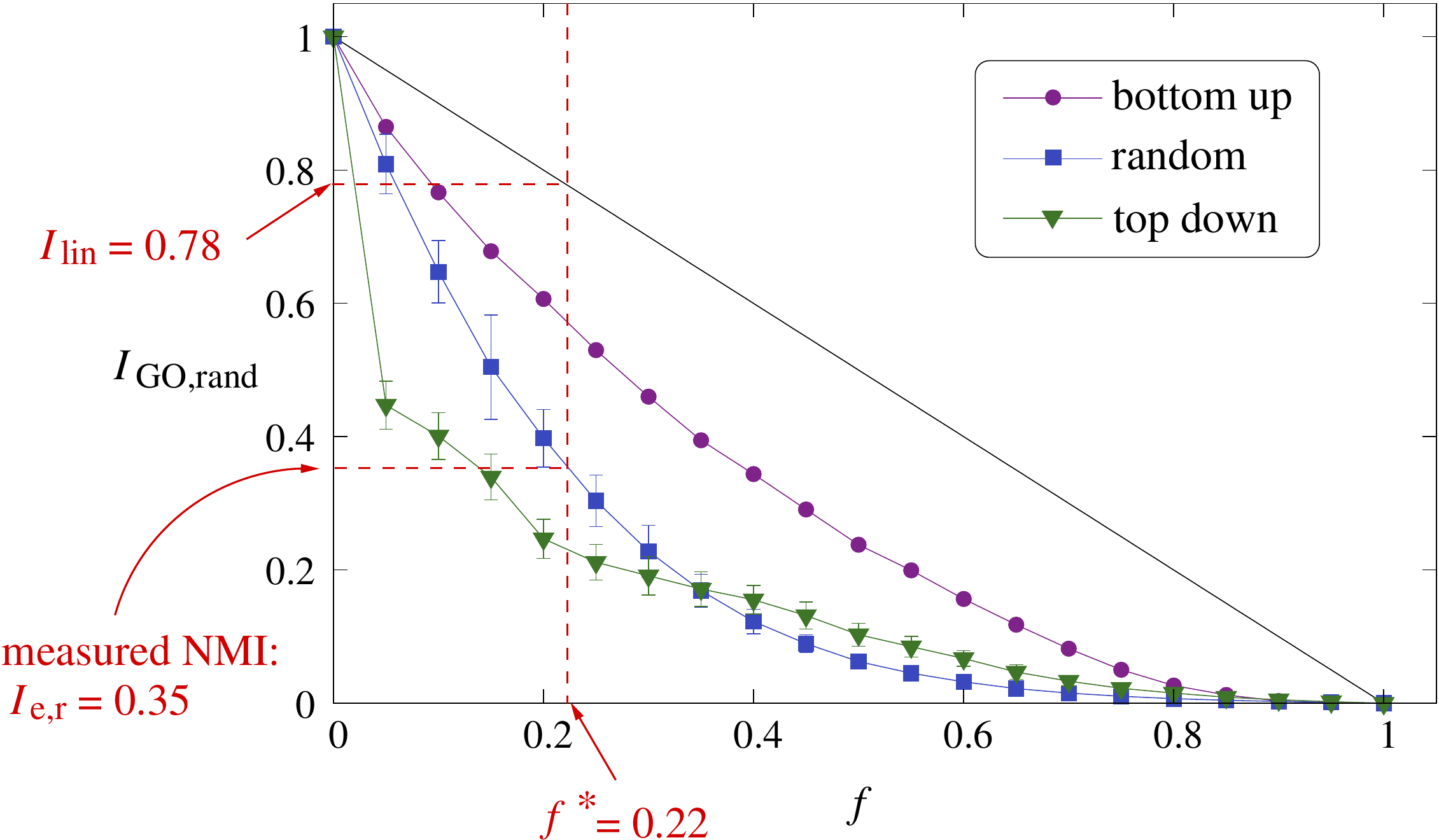}
\end{center}
\begin{flushleft}
\caption{
{\bf NMI decay for the exact hierarchy of protein tags and its randomized counterpart}. We plotted $I$ as given in Eq.(1) of the main paper as a function of the randomly rewired links, $f$. The three different curves correspond to rewiring the links in reverse order according to their position in the hierarchy (purple circles), rewiring in random order (blue squares) and rewiring in the order of the position in the hierarchy (green triangles). The red lines illustrate the calculation of the linearized mutual information for the reconstruction result obtained from algorithm A.
\label{fig:GO_DAG_rand}
}
\end{flushleft}
\end{figure}
 In Fig.\ref{fig:GO_DAG_rand}. the NMI defined in Eq.(1) of the main paper is shown for the exact hierarchy and its randomized counterpart as a function of the randomly rewired links, $f$. The three different curves correspond to three different orders in which the links were chosen for the rewiring: in case of the purple curve we started the rewiring with links pointing to leafs, and continued in reverse order according to the hierarchy, in case of the blue curve, the links were chosen in random order, while in case of the green curve, we started the rewiring at the top of the hierarchy, and continued in the order according to the hierarchy. Similarly to Fig.1. in the main paper, all three curves decay to 0 as $f\rightarrow 1$, thus, the similarity becomes 0 when the compared DAGs become independent. However, the behavior in the small and medium $f$ regime is rather different: the green curve drops below $I_{\rm GO,rand}=0.5$ already at $f=0.05$, while the blue curve shows a moderate decrease and the purple curve decays even more mildly. Similarly to Fig.1. in the main paper, this justifies our statement that the NMI is sensitive also to the position of the links in the hierarchy: rewiring links high in the hierarchy has a larger effect on the similarity compared to rewiring links close to the leafs. Interesting, in the medium $f^*$ regime a crossover can be observed between the green- and the blue curve. The possible explanation for this effect lies in the non-trivial, nor random, nor regular structure of the original DAG.

The red lines in Fig.\ref{fig:GO_DAG_rand}. demonstrate the calculation of the linearized mutual information for the results obtained from algorithm A: The obtained NMI value of $I_{\rm e,r}=0.37$ between the output of the algorithm and the exact DAG is projected to the $f$ axis using the blue curve, resulting in $f^*=0.22$. The linearized mutual information, $I_{\rm lin}$ is given by $1-f^*$, resulting $I_{\rm lin}=0.78$.

\section{Further results on Flickr and IMDb}
\subsection{Additional samples from the Flickr hierarchy}
\begin{figure}[!ht]
\begin{center}
\includegraphics[width=\textwidth]{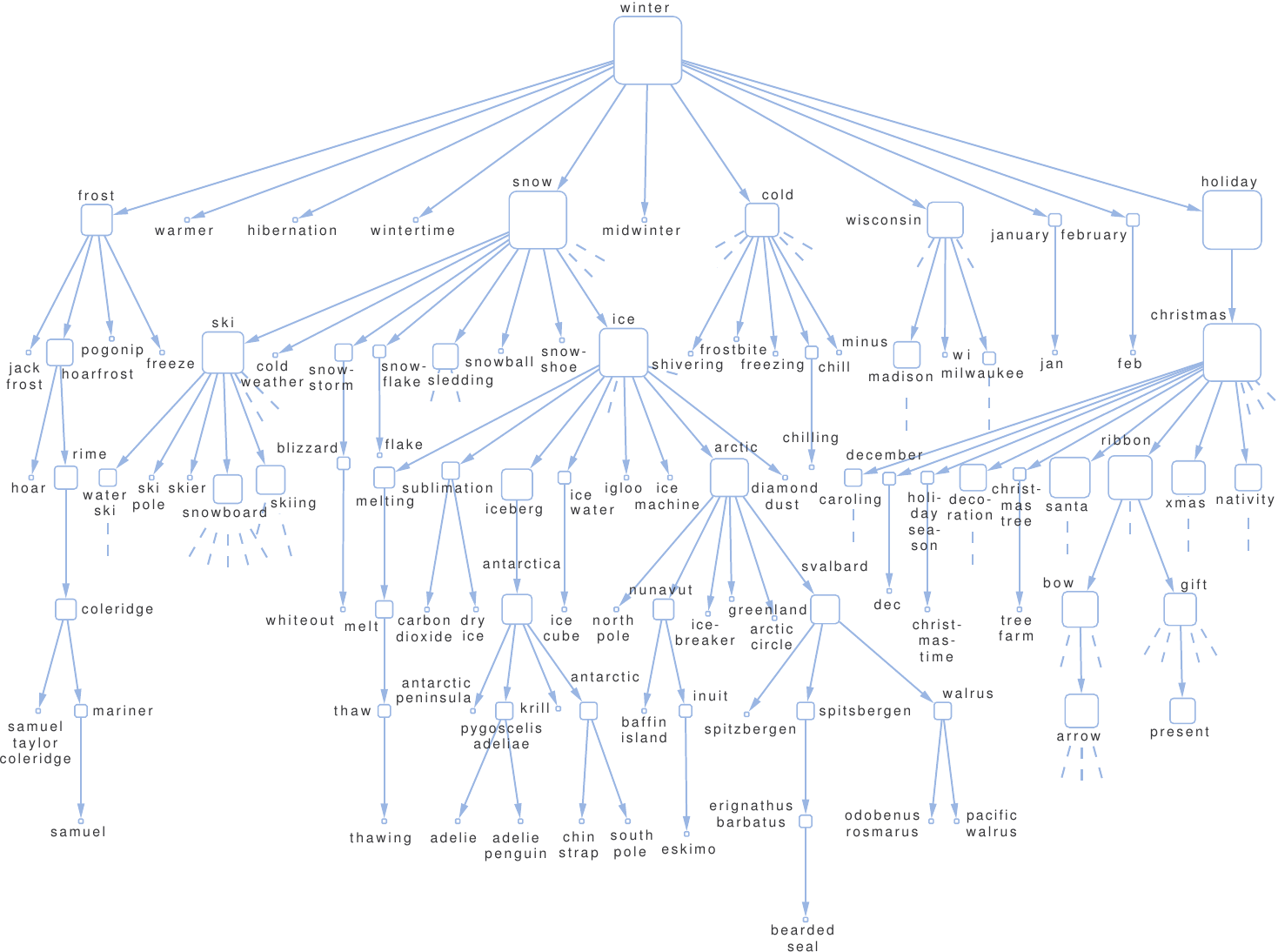}
\end{center}
\begin{flushleft}
\caption{
{\bf Partial subgraph of the descendants of ``winter'' in the hierarchy between Flickr tags obtained from algorithm B.} Stubs (in dashed line) signal further descendants not shown in the figure, and the size of the nodes indicate the total number of descendants.
\label{fig:winter}
}
\end{flushleft}
\end{figure}

\begin{figure}[!ht]
\begin{center}
\includegraphics[width=0.7\textwidth]{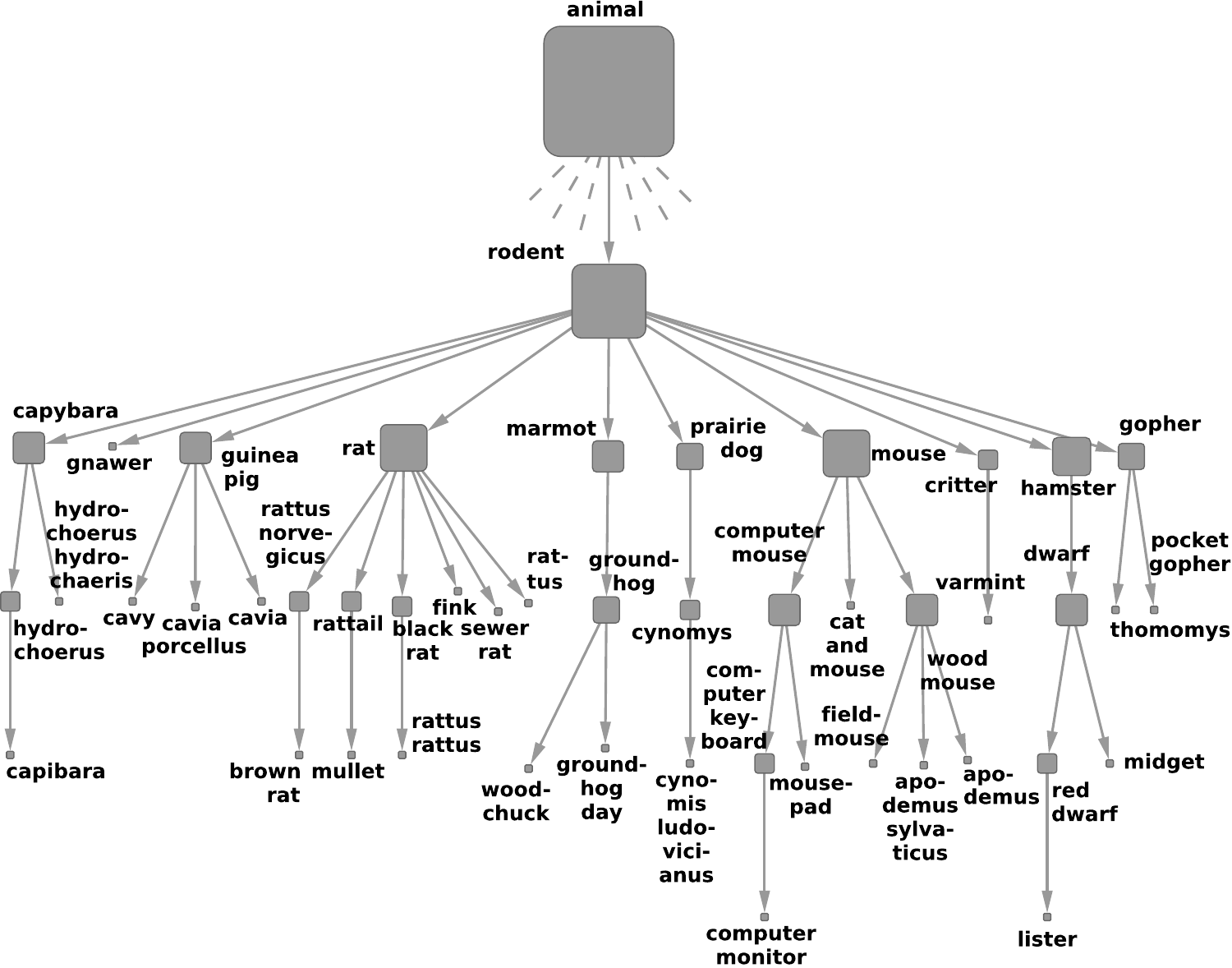}
\end{center}
\begin{flushleft}
\caption{
{\bf A part of the descendants of ``rodent'' in the hierarchy between Flickr tags.} Similarly to Fig.\ref{fig:winter}., the overall hierarchy behind the subgraph shown here was obtained from algorithm B. Stubs (in dashed line) signal further descendants not shown in the figure, and the size of the nodes indicate the total number of descendants.
\label{fig:rodent}
}
\end{flushleft}
\end{figure}

The exact hierarchy between the tags appearing in Flickr is not known, thus, the quality of the extracted hierarchy can be judged only by ``eye'', i.e., by looking at smaller subgraphs, whether they make sense or not. In Fig.3. of the main paper we have already shown a part of the branch under ``reptile'' in the hierarchy obtained from algorithm B. Here we show further examples in the same manner. In Fig.\ref{fig:winter}. we depict a part of the hierarchy under the tag ``winter'', with very reasonable descendants like ``snow'', ``ski'', ``cold'', ``ice'', etc. Similarly, in Fig.\ref{fig:rodent}. we show a part of the descendants of ``rodent'', displaying again a rather meaningful hierarchy.

\subsection{Samples from the hierarchies extracted with the other methods}
\begin{figure}[!ht]
\begin{center}
\includegraphics[width=0.9\textwidth]{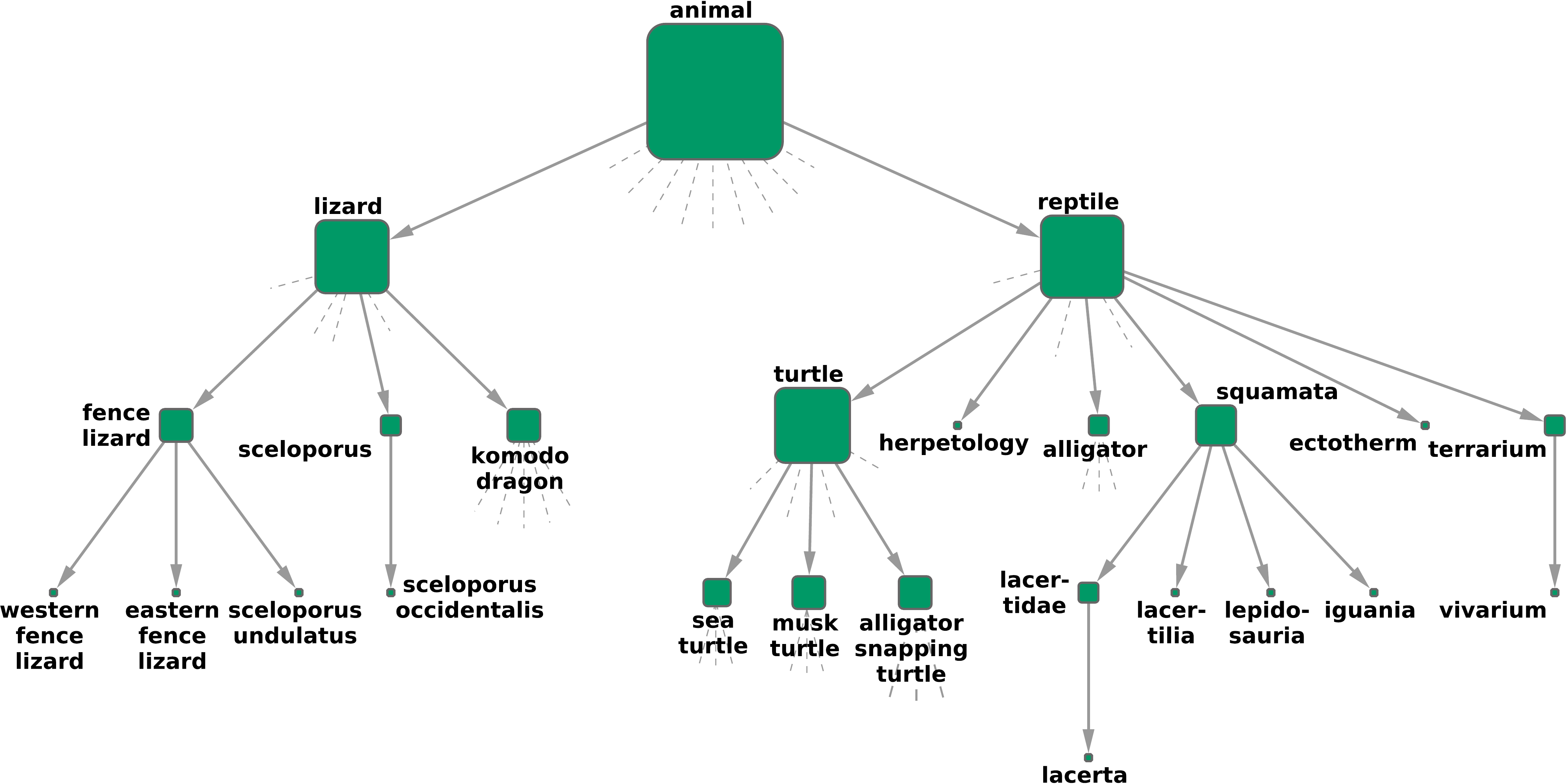}
\end{center}
\begin{flushleft}
\caption{
{\bf A part of the descendants of ``reptile'' and ``lizard'' in the hierarchy between Flickr tags obtained with algorithm A.} Stubs (in dashed line) signal further descendants not shown in the figure, and the size of the nodes indicate the total number of descendants.
\label{fig:reptile_alg_A}
}
\end{flushleft}
\end{figure}

\begin{figure}[!ht]
\begin{center}
\includegraphics[width=\textwidth]{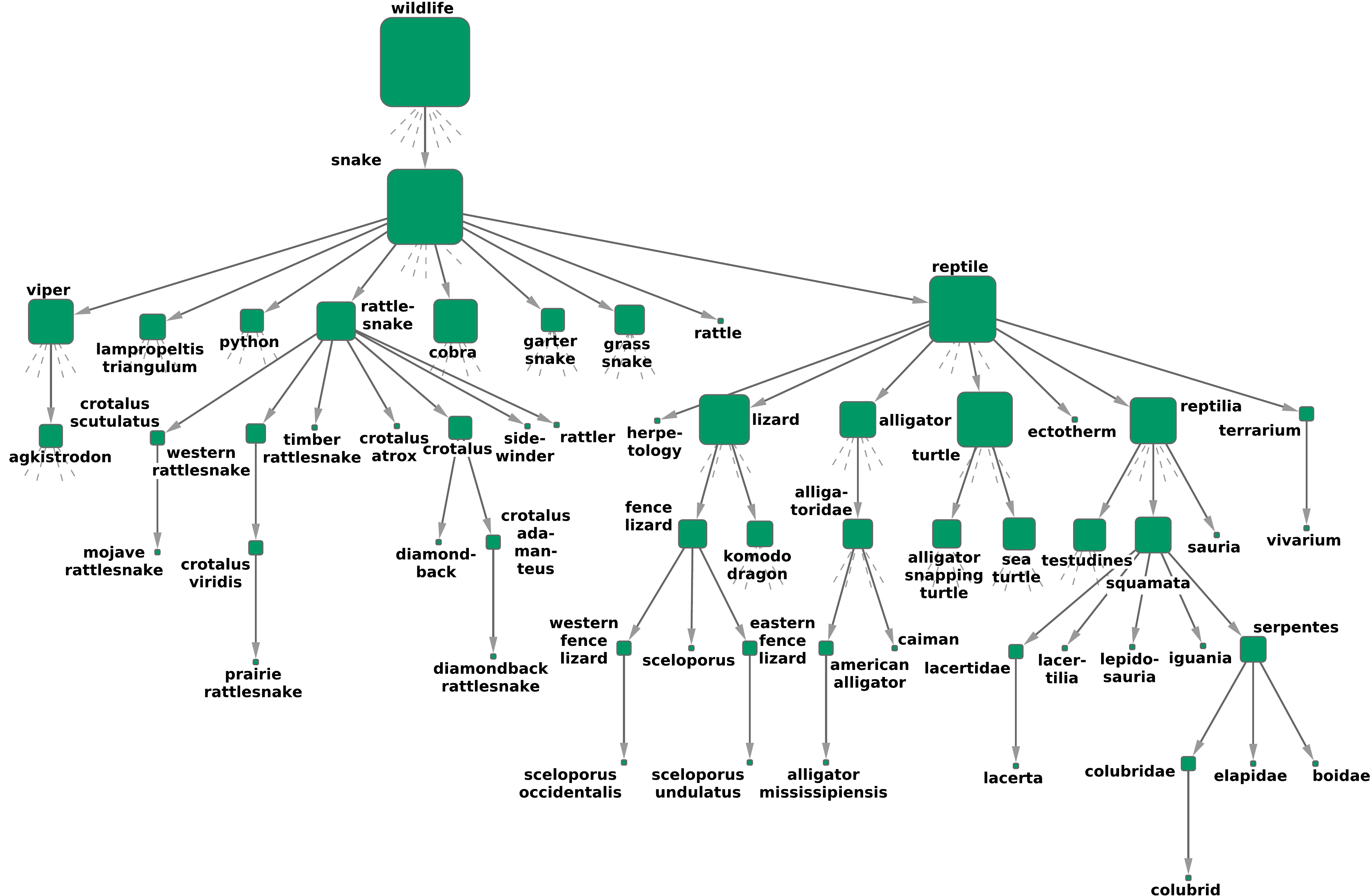}
\end{center}
\begin{flushleft}
\caption{
{\bf A part of the descendants of ``snake'' in the hierarchy between Flickr tags obtained with the method by P.~Heymann \& H.~Garcia-Molina.} Stubs (in dashed line) signal further descendants not shown in the figure, and the size of the nodes indicate the total number of descendants.
\label{fig:reptile_centrality}
}
\end{flushleft}
\end{figure}

\begin{figure}[!ht]
\begin{center}
\includegraphics[width=\textwidth]{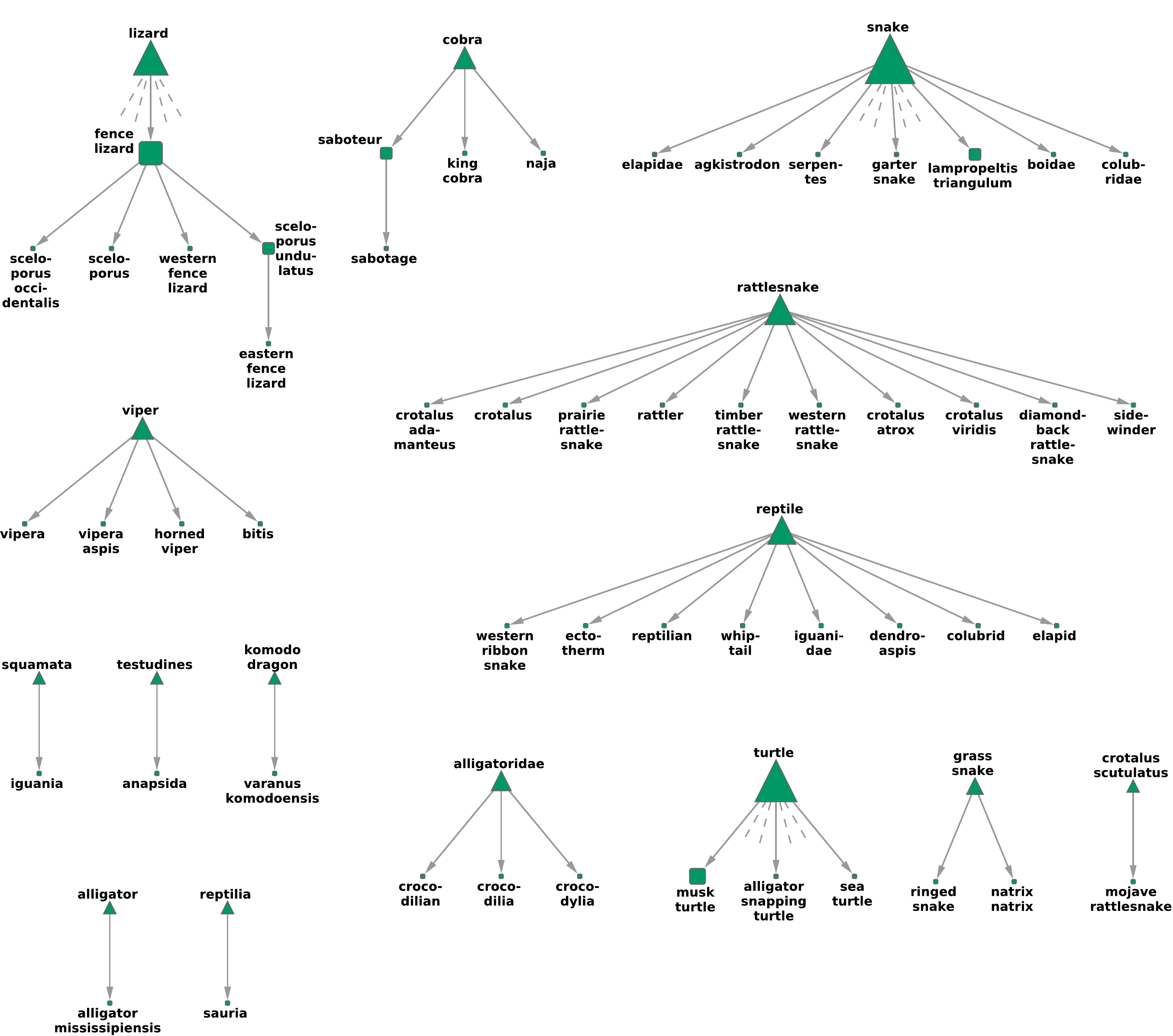}
\end{center}
\begin{flushleft}
\caption{
{\bf Samples from the small hierarchies between Flickr tags obtained with the algorithm by P.~Schmitz.} Triangular shaped nodes represent local roots. These were chosen from tags appearing in Fig.3.\ in the main paper.
\label{fig:reptile_condprob}
}
\end{flushleft}
\end{figure}

\begin{figure}[!ht]
\begin{center}
\includegraphics[width=\textwidth]{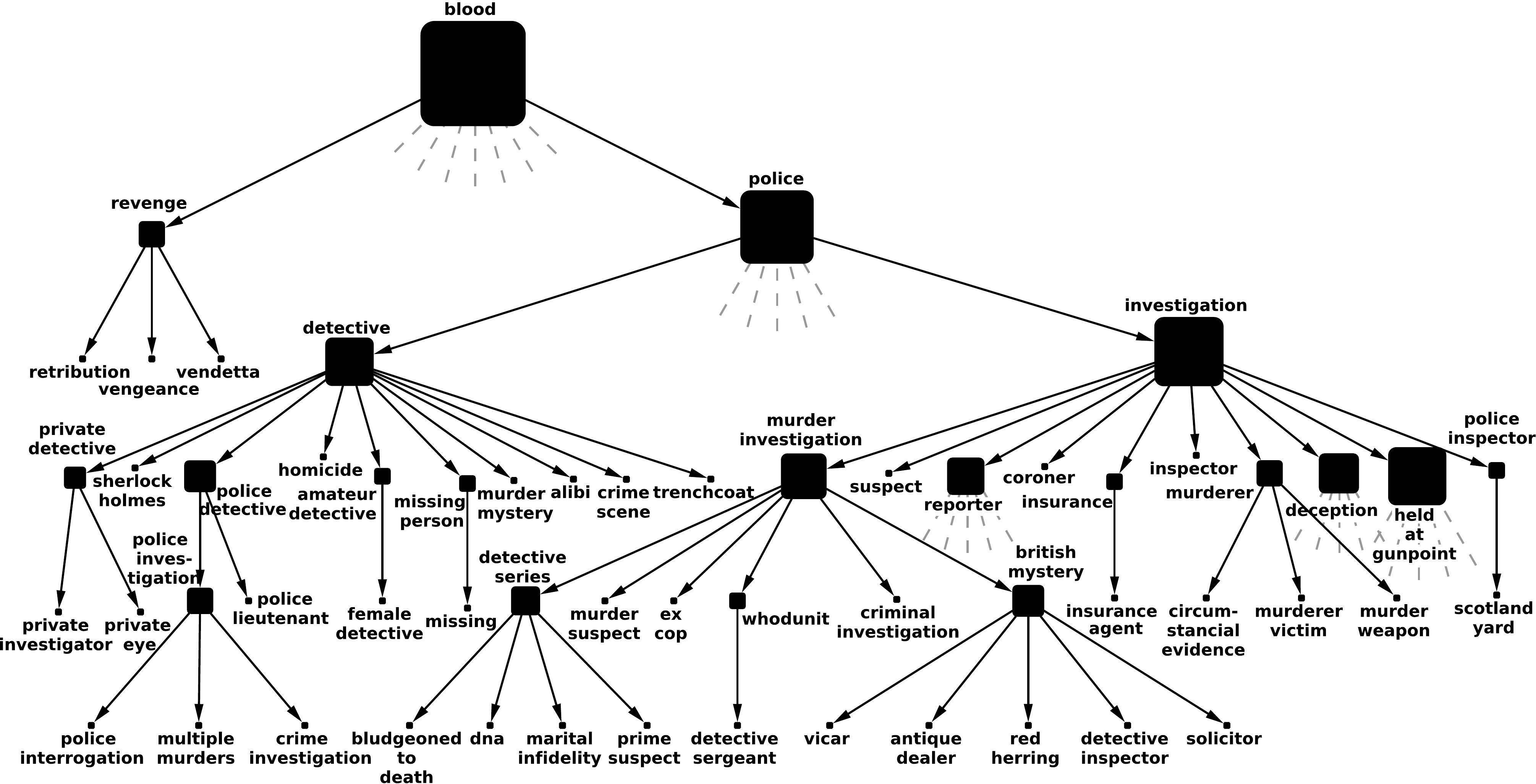}
\end{center}
\begin{flushleft}
\caption{
{\bf A part of the descendants of ``blood'' in the hierarchy between IMDb tags obtained with algorithm A.} Stubs (in dashed line) signal further descendants not shown in the figure, and the size of the nodes indicate the total number of descendants.
\label{fig:murder_alg_A}
}
\end{flushleft}
\end{figure}

\begin{figure}[!ht]
\begin{center}
\includegraphics[width=0.8\textwidth]{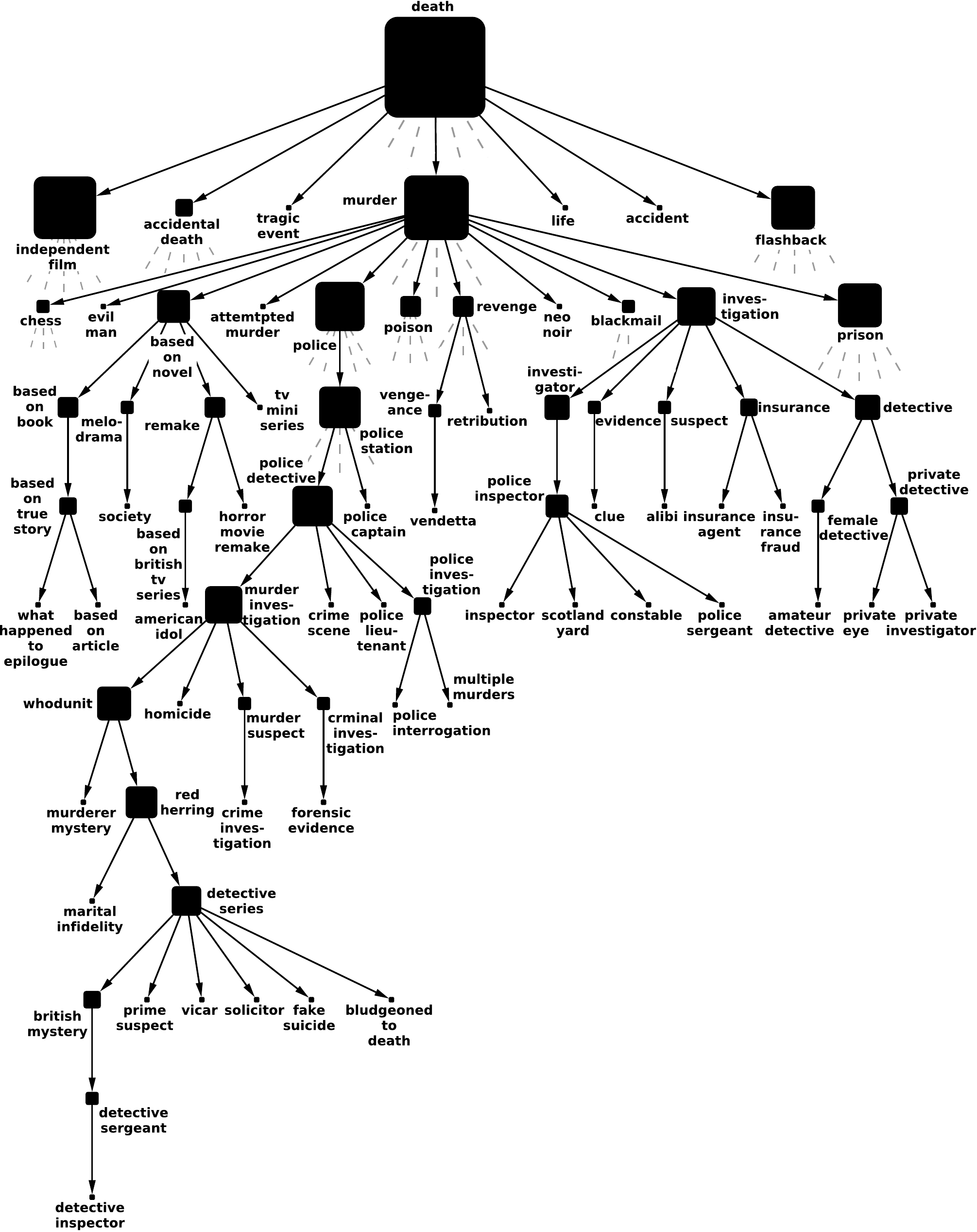}
\end{center}
\begin{flushleft}
\caption{
{\bf A part of the descendants of ``murder'' in the hierarchy between IMDb tags obtained with the method by P.~Heymann \& H.~Garcia-Molina.} Stubs (in dashed line) signal further descendants not shown in the figure, and the size of the nodes indicate the total number of descendants.
\label{fig:murder_centrality}
}
\end{flushleft}
\end{figure}

\begin{figure}[!ht]
\begin{center}
\includegraphics[width=\textwidth]{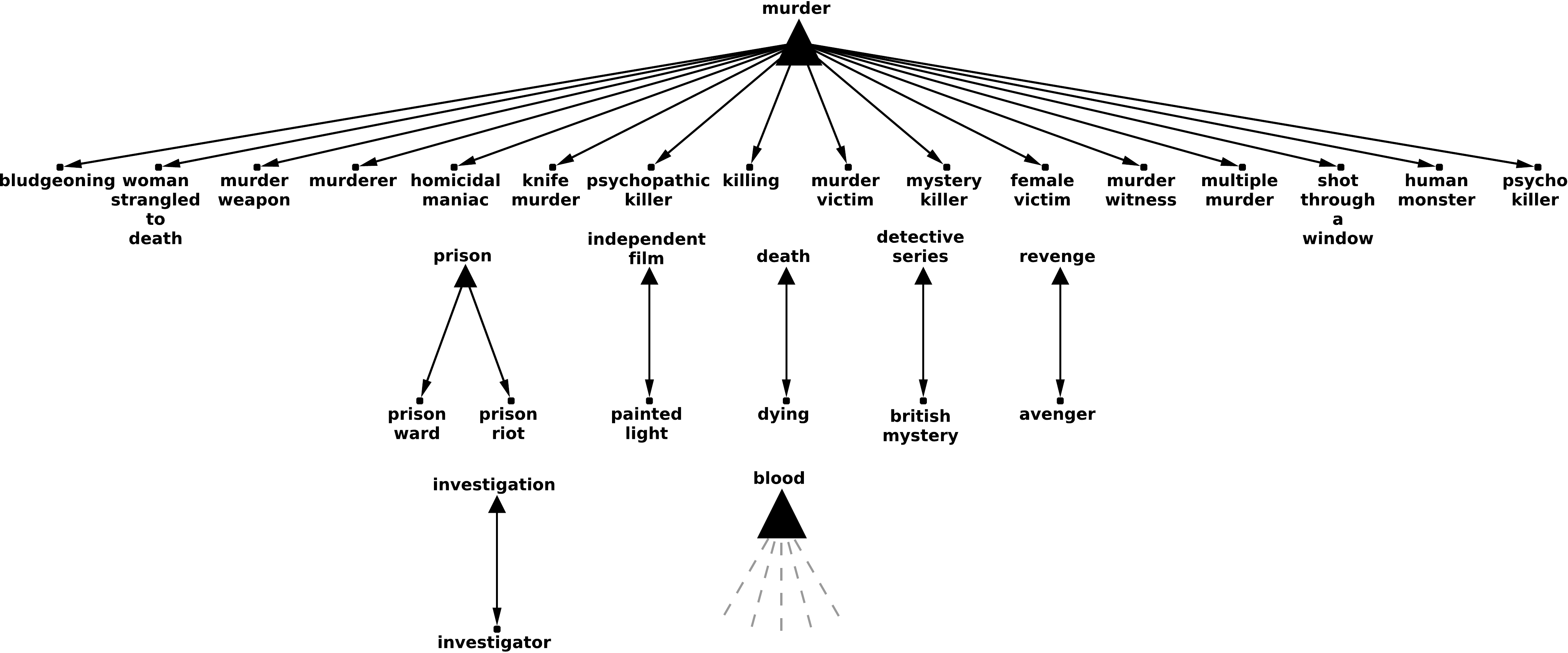}
\end{center}
\begin{flushleft}
\caption{
{\bf Samples from the small hierarchies between IMDb tags obtained with the algorithm by P.~Schmitz} Triangular shaped nodes represent local roots. These were chosen from tags appearing in Fig.4.\ in the main paper.
\label{fig:murder_condprob}
}
\end{flushleft}
\end{figure}

For comparison with Figs.3-4.\ in the main paper, here in Figs.\ref{fig:reptile_alg_A}-\ref{fig:murder_condprob}.\ we show the corresponding parts from the hierarchies extracted with algorithm A, the method by P.~Heymann \& H.~Garcia-Molina and the algorithm by P.~Schmitz. Since the overall structure of the hierarchies is varying over the different algorithms, naturally, the set of tags appearing in these figures is somewhat different compared to Figs.3-4.\ in the main paper. I.e., tags in direct ancestor-descendant relation according to algorithm B can be classified into different branches by an other algorithm or siblings may become unrelated etc. in the output of another method. Therefore, our strategy when preparing Figs.\ref{fig:reptile_alg_A}-\ref{fig:murder_condprob}.\ was to choose the largest branch, containing the most common tags with Figs.3-4.\ in the main paper.

 In Figs.\ref{fig:reptile_alg_A}-\ref{fig:reptile_condprob}.\ we show samples corresponding to Fig.3.\ in the main paper, obtained from the hierarchies extracted for the Flickr tags. Interestingly, in case of algorithm A, (Fig.\ref{fig:reptile_alg_A}.), the tag ``lizard'' and ``reptile'' are classified into different branches. Meanwhile, in the subgraph obtained from the algorithm by P.~Heymann \& H.~Garcia-Molina, (Fig.\ref{fig:reptile_centrality}), the tag ``snake'' has been chosen to be the direct ancestor of ``reptile''. Apart from that, the hierarchy of the tags is rather similar to that shown in Fig.3.\ in the main paper. In case of the algorithm by P.~Schmitz, the obtained result was actually composed of many distinct small hierarchies, with the tags given in Fig.3.\ in the main paper spreading over a large number of different components. Thus, we included a larger set of these small hierarchies in Fig.\ref{fig:reptile_condprob}.\ instead of a single larger subgraph as in Figs.\ref{fig:reptile_alg_A}-\ref{fig:reptile_centrality}. 

In Figs.\ref{fig:murder_alg_A}-\ref{fig:murder_condprob}.\ we show samples from the hierarchies obtained for the IMDb tags. In case of algorithm A, (Fig.\ref{fig:murder_alg_A}.), we display the branch under ``blood'', as most of its descendants appear also on Fig.4 in the main paper, while the tag ``murder'' is missing from the figure, since it was sorted into a different branch. The subgraph shown for the method by P.~Heymann \& H.~Garcia-Molina, (Fig.\ref{fig:murder_centrality}.), has similar features compared to Fig.4 in the main paper, however, the direct ancestor-descendant relation between ``murder'' and ``death'' has been reversed. Finally, the results for the algorithm by P.~Schmitz are again very dispersed, thus, we included more than one small subgraph in Fig.\ref{fig:murder_condprob}.

\section{Synthetic benchmark}

\subsection{Pseudo code}
In Algorithm \ref{benchmark}. we briefly sketch the pseudo code of the preparation of the synthetic tagged data in our benchmark system. As explained in the main paper, the basic idea is to use a random walk process on the pre-defined hierarchy for ensuring the higher frequency of co-occurrences between more closely related tags. Beside the hierarchy between the tags, the following parameters are also assumed to be pre-defined: the number of virtual objects to be generated, the frequency distribution of the tags, the distribution of the number of tags on the objects and the distribution of the random walk lengths.
\begin{algorithm}[!ht]
\caption{Generating synthetic data based on random walk}
\label{benchmark}
\begin{algorithmic}[1]
\ForAll{virtual objects}
  \State draw tag t$_1$ at random according to the tag frequency distribution
  \State assign t$_1$ to the virtual object
  \State draw number of tags n$_{\rm T}$ at random from the distribution of the number of tags on the objects
  \ForAll{i=2, i $<=$ n$_{\rm T}$}
    \If{random number r $< p_{\rm RW}$}
      \State draw random walk length $l_{\rm RW}$ at random from the random walk length distribution
      \State set tag t$_{\rm i}=$t$_1$
      \ForAll{j=1, j $<= l_{\rm RW}$}
        \State random walk on the pre-defined hierarchy, ignoring the link directions:
        \State new tag  t$_{\rm j}:=$ random neighbor of t$_{\rm i}$
        \State set t$_{\rm i}=$t$_{\rm j}$
      \EndFor
      \State assign t$_{\rm i}$ to the virtual object
    \Else
     \State draw tag t$_{\rm i}$ at random according to the tag frequency distribution
     \State assign t$_{\rm i}$ to the virtual object
    \EndIf
  \EndFor
\EndFor
\end{algorithmic}
\end{algorithm}

\subsection{Further tests based on the ``easy'' parameter settings}
\label{sect:test_easy}

\begin{figure}[!ht]
\begin{center}
\includegraphics[width=\textwidth]{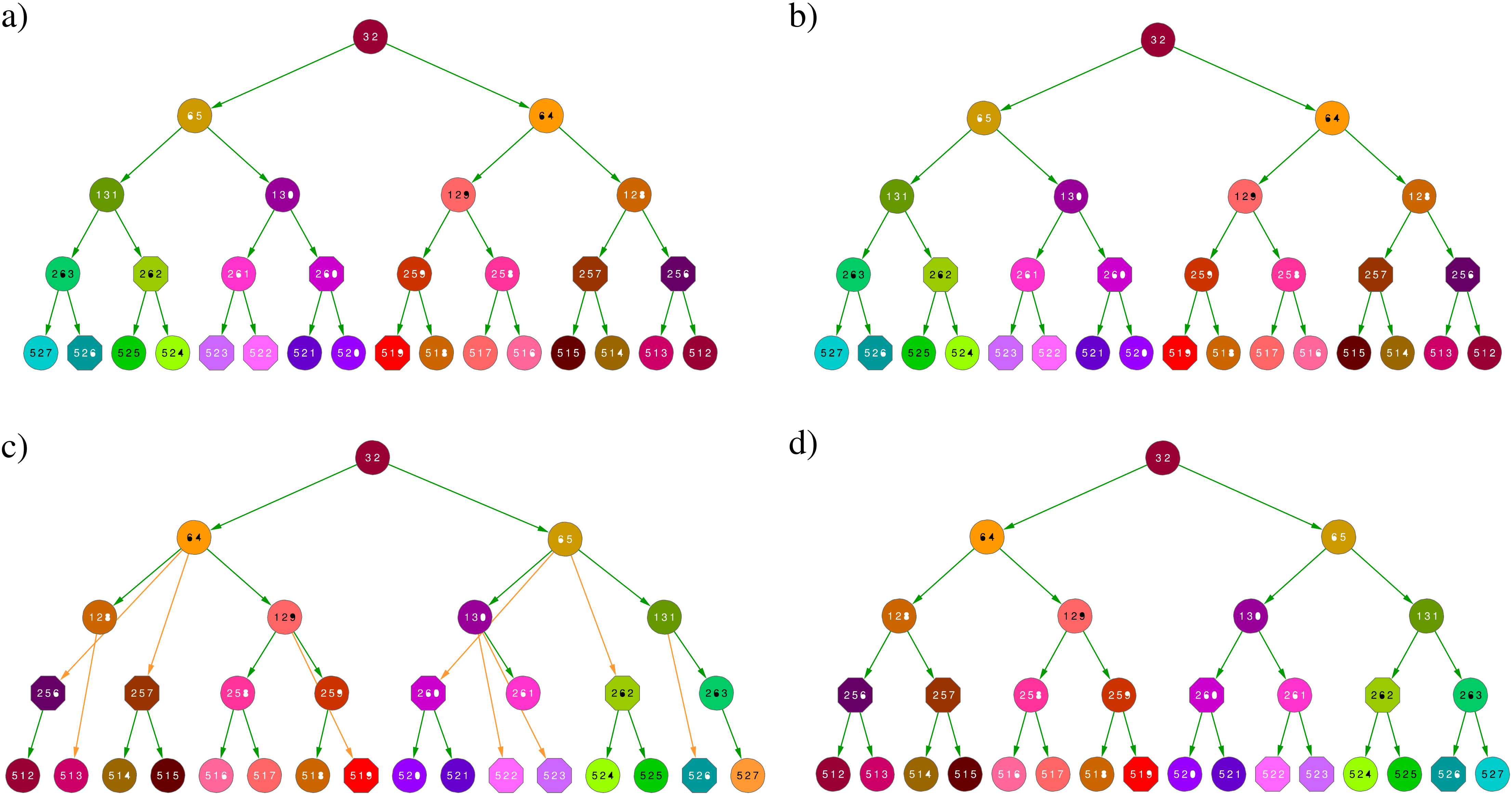}
\end{center}
\begin{flushleft}
\caption{
{\bf Comparison between the exact hierarchy and the reconstructed  hierarchy in case of the ``easy'' computer generated benchmark}. a) A subgraph from the exact hierarchy for testing algorithm A. b) A subgraph from the exact hierarchy for testing algorithm B. c) The subgraph corresponding to a) in the result obtained from algorithm A. Exactly matching links are shown in green, acceptable links are colored orange. d) The subgraph corresponding to b) in the result obtained from algorithm B, showing a perfect match.
\label{fig:easy_bench_result}
}
\end{flushleft}
\end{figure}

In the main paper we have shown that when the frequency of tags is decreasing linearly as a function of the depth in the hierarchy, the synthetic benchmark becomes ``easy'', and an almost perfect reconstruction becomes possible. As an illustration, in Fig.\ref{fig:easy_bench_result}a-b we show parts from the exact DAGs, (binary trees of 1023 tags), used for testing algorithm A and algorithm B, respectively. In Fig.\ref{fig:easy_bench_result}c we display the corresponding subgraph from the hierarchy obtained from algorithm A. The result is quite good, where the majority of the links are exactly matching, (colored green), while the rest are acceptable (shown in orange). However in case of algorithm B the chosen part of the reconstruction is perfect, as shown in Fig.\ref{fig:easy_bench_result}d, with only exactly matching links.

According to the results discussed in the main paper, when the tag frequencies are independent of the position in the hierarchy and have a power-law distribution, the benchmark becomes hard. Here we examine the effect of changes in the other parameters of the benchmark. First, our starting point is the ``easy'' parameter setting, while the results obtained for the ``hard'' parameter setting are discussed in Sect.\ref{sect:test_hard}. As mentioned in the main paper, the most important feature of the ``easy'' parameter settings is that the frequency of the tags is decreasing linearly as a function of the level depth in the hierarchy. The other parameters were set as follows: an average number of 3 co-occurring tags were generated on altogether 2,000,000 hypothetical objects, with random walk probability of $p_{\rm RW}=0.5$ and random walk lengths chosen from a uniform distribution between 1 and 3, (the results are shown in Table 2. in the main paper). First we study the effect of changing the length of the random walks. 
\begin{table}[!ht]
\caption{
\bf{Quality measures of the reconstructed hierarchies with random walk length of 1 step.}}
\begin{tabular}{|l|c|c|c|c|c|c|c|}
\hline
    & $r_{\rm E}$ & $r_{\rm A}$ & $r_{\rm I}$ & $r_{\rm U}$ & $r_{\rm M}$ & $I_{\rm e,r}$ & $I_{\rm lin}$
 \\ \hline  
algorithm A & 68\%   &95\% & 0\% & 5\%& 0\% & 47\% & 86\% \\
\hline
algorithm B &  100\% & 100\% & 0\% & 0\% & 0\% & 100\% & 100\% \\
\hline 
P.~Heymann \& H.~Garcia-Molina & 99\% & 99\% & 1\% & 0\%&  0\%& 92\% & 99\% \\
\hline
P.~Schmitz &  0\% & 0\% & 0\% & 0\% & 100\% & 0\% & 0\% \\
\hline
\end{tabular}
\begin{flushleft}
The setting of the other parameters were exactly the same as in case of the ``easy'' synthetic data set discussed in the main paper.
\end{flushleft}
\label{table:rw_step_1}
\end{table}
In Table \ref{table:rw_step_1}. we show the results when we decrease the random walk length to only a single step: according to the listed measures, the quality of the reconstruction for Algorithm B, the method by  P.~Heymann \& H.~Garcia-Molina and the algorithm by P.~Schmitz remain exactly or almost exactly the same. In case of Algorithm A the quality indicators are somewhat lower compared to Table 2. in the main text, however, solely $I_{\rm e,r}$ is changed significantly.
\begin{table}[!ht]
\caption{
\bf{Quality measures of the reconstructed hierarchies with maximum random walk length of 5 step.}}
\begin{tabular}{|l|c|c|c|c|c|c|c|}
\hline
    & $r_{\rm E}$ & $r_{\rm A}$ & $r_{\rm I}$ & $r_{\rm U}$ & $r_{\rm M}$ & $I_{\rm e,r}$ & $I_{\rm lin}$
 \\ \hline  
algorithm A & 95\%   &100\% & 0\% & 0\%& 0\% & 99\% & 100\% \\
\hline
algorithm B &  100\% & 100\% & 0\% & 0\% & 0\% & 100\% & 100\% \\
\hline 
P.~Heymann \& H.~Garcia-Molina & 99\% & 99\% & 0\% & 0\%&  0\%& 93\% & 99\% \\
\hline
P.~Schmitz &  0\% & 0\% & 0\% & 0\% & 100\% & 0\% & 0\% \\
\hline
\end{tabular}
\begin{flushleft}
The setting of the other parameters were exactly the same as in case of the ``easy'' synthetic data set discussed in the main paper.
\end{flushleft}
\label{table:rw_step_5}
\end{table}
In Table \ref{table:rw_step_5}. we show the results when the length of the random walks was chosen from a uniform distribution between 1 and 5, and the other parameters of the data set were left the same. Again, algorithm B, the method by  P.~Heymann \& H.~Garcia-Molina and the algorithm by P.~Schmitz produce the same (or almost the same) results as presented in Table 2. of the main text. The results from algorithm A are now better compared to the original settings, reaching almost the same quality as algorithm A. In conclusion, the change in the length of the random walks has only a negligible effect for three out of the four methods studied here, and a mild effect on the results from the fourth one.

Next, we examine the effect of reducing the number of generated virtual objects. 
\begin{table}[!ht]
\caption{
\bf{Quality measures of the reconstructed hierarchies with 200,000 hypothetical objects}}
\begin{tabular}{|l|c|c|c|c|c|c|c|}
\hline
    & $r_{\rm E}$ & $r_{\rm A}$ & $r_{\rm I}$ & $r_{\rm U}$ & $r_{\rm M}$ & $I_{\rm e,r}$ & $I_{\rm lin}$
 \\ \hline  
algorithm A & 67\%   &97\% & 1\% & 2\%& 0\% & 86\% & 98\% \\
\hline
algorithm B &  98\% & 98\% & 2\% & 0\% & 0\% & 100\% & 100\% \\
\hline 
P.~Heymann \& H.~Garcia-Molina & 98\% & 98\% & 2\% & 0\%&  0\%& 93\% &99\% \\
\hline
P.~Schmitz &  0\% & 0\% & 0\% & 0\% & 100\% & 0\% & 0\% \\
\hline
\end{tabular}
\begin{flushleft}
The setting of the other parameters were exactly the same as in case of the ``easy'' synthetic data set discussed in the main paper.
\end{flushleft}
\label{table:200_obj}
\end{table}
In Table \ref{table:200_obj}. we show the results obtained when we generated only 200,000 virtual objects instead of 2,000,000, (and otherwise used the same parameters as in case of the ``easy'' synthetic data set). Not surprisingly, the quality of the methods show a slight decrease, as the hierarchy has to be reconstructed based on less information. However, the effect is only minor. 
\begin{table}[!ht]
\caption{
\bf{Quality measures of the reconstructed hierarchies with 50,000 hypothetical objects}}
\begin{tabular}{|l|c|c|c|c|c|c|c|}
\hline
    & $r_{\rm E}$ & $r_{\rm A}$ & $r_{\rm I}$ & $r_{\rm U}$ & $r_{\rm M}$ & $I_{\rm e,r}$ & $I_{\rm lin}$
 \\ \hline  
algorithm A & 57\%   &89\% & 3\% & 8\%& 0\% & 75\% & 95\% \\
\hline
algorithm B &  70\% & 81\% & 13\% & 6\% & 0\% & 76\% & 95\% \\
\hline 
P.~Heymann \& H.~Garcia-Molina & 87\% & 91\% & 5\% & 4\%&  0\%& 94\% &99\% \\
\hline
P.~Schmitz &  2\% & 3\% & 0\% & 0\% & 97\% & 1\% & 11\% \\
\hline
\end{tabular}
\begin{flushleft}
The setting of the other parameters were exactly the same as in case of the ``easy'' synthetic data set discussed in the main paper.
\end{flushleft}
\label{table:50_obj}
\end{table}
When reducing the number of objects further down to 50,000, the drop in the quality measures becomes more pronounced, as presented in Table \ref{table:50_obj}. Interestingly, the algorithm by P.~Schmitz shows a different behavior, with a slight increase in quality. As mentioned in the main paper, the study of the reasons for the outlying behavior of this algorithm on the synthetic data is out of the scope of present work.

\begin{table}[!ht]
\caption{
\bf{Quality measures of the reconstructed hierarchies with random walk probability $p_{\rm RW}=0.1$.}}
\begin{tabular}{|l|c|c|c|c|c|c|c|}
\hline
    & $r_{\rm E}$ & $r_{\rm A}$ & $r_{\rm I}$ & $r_{\rm U}$ & $r_{\rm M}$ & $I_{\rm e,r}$ & $I_{\rm lin}$
 \\ \hline  
algorithm A & 1\%   &61\% & 0\% & 39\%& 0\% & 0\% & 1\% \\
\hline
algorithm B &  89\% & 90\% & 8\% & 2\% & 0\% & 64\% & 92\% \\
\hline 
P.~Heymann \& H.~Garcia-Molina & 88\% & 99\% & 1\% & 0\%&  0\%& 68\% &93\% \\
\hline
P.~Schmitz &  0\% & 0\% & 0\% & 0\% & 100\% & 0\% & 0\% \\
\hline
\end{tabular}
\begin{flushleft}
The setting of the other parameters were exactly the same as in case of the ``easy'' synthetic data set discussed in the main paper.
\end{flushleft}
\label{table:rw_prob}
\end{table}
Finally, in Table \ref{table:rw_prob}. we show the quality measures obtained when the random walk probability was reduced from $p_{\rm RW}=0.5$ to $p_{\rm RW}=0.1$, (and the other parameters were the same as in case of the ``easy'' synthetic data set). Similarly to the case of reducing the number of objects, this provides a more difficult task for the tag hierarchy extracting algorithms, as most of the tags are chosen at random on the objects. Accordingly, the quality measures are decreased when compared to the results shown in Table 2. of the main text. This effect is quite significant in case of algorithm A, while its less pronounced for algorithm B and the method by P.~Heymann and H.~Garcia-Molina.

\subsection{Further tests based on the ``hard'' parameter settings}
\label{sect:test_hard}
In similar fashion to Sect.\ref{sect:test_easy}, here we examine the effects of changing the parameters when we start from the ``hard'' parameter setting. As mentioned in the main paper, the main feature making this choice of parameters ``hard'' is that the frequency of tags is independent of the level depth in the hierarchy. Otherwise, the parameters of the data set discussed in Table 3.\ of the main text were the following: an average number of 3 co-occurring tags were generated on altogether 2,000,000 hypothetical objects, with random walk probability of $p_{\rm RW}=0.5$ and random walk lengths chosen from a uniform distribution between 1 and 3. Starting from this parameter setting, in Table \ref{table:rw_step_1_hard}.\ we show the results obtained when the random walk length is reduced to 1. For all 4 methods, we can observe a slight increase in the quality, however, no significant changes have occurred when comparing to Table 3.\ in the main text.
\begin{table}[!ht]
\caption{
\bf{Quality measures of the reconstructed hierarchies with random walk length of 1 step.}}
\begin{tabular}{|l|c|c|c|c|c|c|c|}
\hline
    & $r_{\rm E}$ & $r_{\rm A}$ & $r_{\rm I}$ & $r_{\rm U}$ & $r_{\rm M}$ & $I_{\rm e,r}$ & $I_{\rm lin}$
 \\ \hline  
algorithm A & 40\%   & 40\% & 17\% & 43\%& 0\% & 21\% & 70\% \\
\hline
algorithm B &  92\% & 93\% & 5\% & 2\% & 0\% & 84\% & 97\% \\
\hline 
P.~Heymann \& H.~Garcia-Molina & 51\% & 55\% & 30\% & 15\%&  0\%& 28\% & 76\% \\
\hline
P.~Schmitz &  4\% & 4\% & 0\% & 5\% & 91\% & 2\% & 18\% \\
\hline
\end{tabular}
\begin{flushleft}
The setting of the other parameters were exactly the same as in case of the ``hard'' synthetic data discussed in the main paper.
\end{flushleft}
\label{table:rw_step_1_hard}
\end{table}

In Table \ref{table:rw_step_5_hard}.\ we show the results when the length of the random walks was chosen from a uniform distribution between 1 and 5, and the other parameters of the data set were left the same as in case of Table 3.\ in the main text. Interestingly, this time the quality measures have been lowered slightly, nevertheless, no significant change can be observed. In a similar fashion to Sect.\ref{sect:test_easy}, our conclusion is that the length of the random walk has no significant effect on the quality of the examined algorithms.
\begin{table}[!ht]
\caption{
\bf{ Quality measures of the reconstructed hierarchies with maximum random walk length of 5 step.}}
\begin{tabular}{|l|c|c|c|c|c|c|c|}
\hline
    & $r_{\rm E}$ & $r_{\rm A}$ & $r_{\rm I}$ & $r_{\rm U}$ & $r_{\rm M}$ & $I_{\rm e,r}$ & $I_{\rm lin}$
 \\ \hline  
algorithm A & 28\%   &36\% & 29\% & 35\%& 0\% & 18\% & 66\% \\
\hline
algorithm B &  85\% & 88\% & 10\% & 2\% & 0\% & 81\% & 96\% \\
\hline 
P.~Heymann \& H.~Garcia-Molina & 46\% & 52\% & 34\% & 14\%&  0\%& 28\% & 76\% \\
\hline
P.~Schmitz &  1\% & 1\% & 1\% & 4\% & 94\% & 1\% & 4\% \\
\hline
\end{tabular}
\begin{flushleft}
The setting of the other parameters were exactly the same as in case of the ``hard'' synthetic data set discussed in the main paper.
\end{flushleft}
\label{table:rw_step_5_hard}
\end{table}

We continue our experiments by changing the number of virtual objects in the preparation of the data set. In Table \ref{table:200_obj_hard}.\ we show the results obtained when we generated only 200,000 virtual objects instead of 2,000,000, (and otherwise used the same parameters as in case of the ``hard'' synthetic data set). The quality measures for algorithm A, the method by P. Heymann \& H. Garcia-Molina and the algorithm by P. Schmitz remained almost the same, while the marks for algorithm B have been slightly reduced, (however, algorithm B is still far the best method on this data set).
\begin{table}[!ht]
\caption{
\bf{Quality measures of the reconstructed hierarchies with 200,000 hypothetical objects}}
\begin{tabular}{|l|c|c|c|c|c|c|c|}
\hline
    & $r_{\rm E}$ & $r_{\rm A}$ & $r_{\rm I}$ & $r_{\rm U}$ & $r_{\rm M}$ & $I_{\rm e,r}$ & $I_{\rm lin}$
 \\ \hline  
algorithm A & 31\%   &36\% & 26\% & 38\%& 0\% & 18\% & 66\% \\
\hline
algorithm B &  80\% & 86\% & 12\% & 2\% & 0\% & 76\% & 95\% \\
\hline 
P.~Heymann \& H.~Garcia-Molina & 48\% & 54\% & 32\% & 14\%&  0\%& 29\% & 76\% \\
\hline
P.~Schmitz &  1\% & 2\% & 0\% & 4\% & 94\% & 1\% & 5\% \\
\hline
\end{tabular}
\begin{flushleft}
The setting of the other parameters were exactly the same as in case of the ``hard'' synthetic data set discussed in the main paper.
\end{flushleft}
\label{table:200_obj_hard}
\end{table}
In Table \ref{table:50_obj_hard}.\ we show the results obtained when the number of hypothetical objects were further reduced to 50,000. In this case the quality of algorithm A, the method by P. Heymann \& H. Garcia-Molina and the algorithm by P. Schmitz has slightly dropped, when compared to Table 3.\ in the main paper. The decrease in the quality is more pronounced in case of algorithm B, however, its results are still much better than that of the others. In conclusion, the lowering of the number of virtual objects affects most the result from algorithm B, nevertheless its quality was always significantly higher compared to the other methods.
\begin{table}[!ht]
\caption{
\bf{Quality measures of the reconstructed hierarchies with 50,000 hypothetical objects}}
\begin{tabular}{|l|c|c|c|c|c|c|c|}
\hline
    & $r_{\rm E}$ & $r_{\rm A}$ & $r_{\rm I}$ & $r_{\rm U}$ & $r_{\rm M}$ & $I_{\rm e,r}$ & $I_{\rm lin}$
 \\ \hline  
algorithm A & 29\%   &36\% & 26\% & 39\%& 0\% & 17\% & 65\% \\
\hline
algorithm B &  66\% & 74\% & 20\% & 6\% & 0\% & 55\% & 89\% \\
\hline 
P.~Heymann \& H.~Garcia-Molina & 46\% & 53\% & 33\% & 14\%&  0\%& 28\% & 76\% \\
\hline
P.~Schmitz &  1\% & 2\% & 0\% & 5\% & 93\% & 1\% & 6\% \\
\hline
\end{tabular}
\begin{flushleft}
The setting of the other parameters were exactly the same as in case of the ``hard'' synthetic data set discussed in the main paper.
\end{flushleft}
\label{table:50_obj_hard}
\end{table}

Finally, in Table \ref{table:rw_prob_hard}.\ we examine the effects of lowering the random walk probability from $p_{\rm RW}=0.5$ to $p_{\rm RW}=0.1$,(while keeping the other parameters the same as in case of the ``hard'' synthetic data set). As mentioned in Sect.\ref{sect:test_easy}, this provides a more difficult task for the tag hierarchy extracting algorithms, as most of the tags are chosen at random on the objects. Accordingly, the quality measures are decreased when compared to the results shown in Table 3.\ of the main text. However, this effect is quite significant in case of algorithm A, while it is more mild for the method by P.\ Heymann \& H.\ Garcia-Molina, and is even less pronounced in case of algorithm B.
\begin{table}[!ht]
\caption{
\bf{Quality measures of the reconstructed hierarchies with random walk probability $p_{\rm RW}=0.1$.}}
\begin{tabular}{|l|c|c|c|c|c|c|c|}
\hline
    & $r_{\rm E}$ & $r_{\rm A}$ & $r_{\rm I}$ & $r_{\rm U}$ & $r_{\rm M}$ & $I_{\rm e,r}$ & $I_{\rm lin}$
 \\ \hline  
algorithm A & 10\%   &12\% & 14\% & 74\%& 0\% & 5\% & 33\% \\
\hline
algorithm B &  65\% & 71\% & 21\% & 8\% & 0\% & 61\% & 91\% \\
\hline 
P.~Heymann \& H.~Garcia-Molina & 35\% & 36\% & 34\% & 30\%&  0\%& 18\% &66\% \\
\hline
P.~Schmitz &  0\% & 0\% & 0\% & 7\% & 93\% & 0\% & 0\% \\
\hline
\end{tabular}
\begin{flushleft}
The setting of the other parameters were exactly the same as in case of the ``hard'' synthetic data set discussed in the main paper.
\end{flushleft}
\label{table:rw_prob_hard}
\end{table}
Our general conclusion regarding the robustness of the examined algorithms is that no significant differences could be observed in our experiments on the synthetic data sets. Algorithm B showed more sensitivity to the number of virtual objects compared to algorithm A and the method by P.~Heymann \& H.~Garcia-Molina. In contrast, when reducing the random walk probability, algorithm B was found to be more robust compared to the other methods.